\documentclass[a4paper]{elsarticle}
\usepackage{graphicx,amssymb,color,url,amscd,stmaryrd,supertabular,amsmath,tikz,cancel}
\usetikzlibrary{shapes,snakes,arrows,automata,calc}

\def\qed{\relax\ifmmode\hskip2em \blacksquare\else\unskip\nobreak\hfill\hskip1em $\blacksquare$\fi}
\def\exqed{\relax\ifmmode\hskip2em \diamond\else\unskip\nobreak\hfill\hskip1em $\diamond$\fi}

\newcommand{\ZZ}{\mathbb{Z}}
\newcommand{\NN}{\mathbb{N}}

\newcommand\card[1]{\left|{#1}\right|}

\newcommand{\ACA}{\mathcal{A}}
\newcommand{\ACB}{\mathcal{B}}
\newcommand{\ACC}{\mathcal{C}}

\newcommand{\ACX}{\mathcal{X}}
\newcommand{\ACU}{\mathcal{U}}
\newcommand{\locA}{f_{\ACA}}
\newcommand{\locB}{f_{\ACB}}
\newcommand{\locC}{f_{\ACC}}
\newcommand\glob[1]{G_{#1}}
\newcommand\local[1]{f_{#1}}
\newcommand{\globA}{G_{\ACA}}
\newcommand{\globB}{G_{\ACB}}
\newcommand{\globC}{G_{\ACC}}
\newcommand{\globX}{G_{\ACX}}
\newcommand\alphabe[1]{S_{#1}}
\newcommand{\alphA}{\alphabe{\ACA}}
\newcommand{\alphB}{\alphabe{\ACB}}
\newcommand{\alphC}{\alphabe{\ACC}}
\newcommand{\radA}{r_{\ACA}}
\newcommand{\radB}{r_{\ACB}}

\newcommand\sfbop[1]{\operatorname{\hbox{\usefont{T1}{cmss}{bx}{n}#1}}}

\newcommand\opdiv{\sfbop{divide}}

\newcommand\tPmCS{\sfbop{\~{P}CS}}

\newcommand{\unif}{\overline}
\newcommand\sac{\sqsubseteq}
\newcommand{\sacby}[1]{\sqsubseteq_{#1}}
\newcommand{\fac}{\trianglelefteq}
\newcommand{\facby}[1]{\trianglelefteq_{#1}}
\newcommand{\facsac}{\fac\!\sac}
\newcommand\bulk[2]{{#1}^{\left\langle{#2}\right\rangle}}
\newcommand{\simu}{\preccurlyeq}
\newcommand{\sacsimu}{\simu_i}
\newcommand{\facsimu}{\simu_s}
\newcommand{\facsacsimu}{\simu_m}
\newcommand{\sime}{\sim}
\newcommand{\sacsime}{\sime_i}

\newcommand{\simc}[1]{\left[{#1}\right]}
\newcommand{\sacsimc}[1]{\left[{#1}\right]_i}

\newcommand{\isom}{\equiv}
\newcommand{\locrelset}{\mathcal{R}}

\newcommand\grporder{\leqslant_\square}
\newcommand\grp[2]{{#1}^{[#2]}}
\newcommand\grpclass[1]{\left[{#1}\right]_{\Box}}

\newcommand\MTM{\mathcal{M}}
\newcommand\statM{S_\MTM}
\newcommand\rubM{Q_\MTM}
\newcommand\transM{\phi_\MTM}

\newcommand\dist{d}

\newcommand\sing{\bot}
\newcommand\univ{\mathfrak{U}}
\newcommand\zee[1]{\mathcal{Z}_{#1}}
\newcommand\shi[2]{\sigma_{{#1},{#2}}}
\newcommand\carashi[1]{\chi\left({#1}\right)}

\newcommand{\limprod}[1]{{\ACA}_\infty}
\newcommand\blc[1]{\texttt{B}_{#1}}
\newcommand\flagseq[1]{u_{#1}}

\newcommand\equi{\mathcal{T}_1}
\newcommand\equipt{\mathcal{T}_2}
\newcommand\sensi{\mathcal{T}_3}
\newcommand\expansi{\mathcal{T}_4}
\newcommand\fami{\mathfrak{F}}
\newcommand\cod{\phi}

\newtheorem{defn}{Definition}[section]
\newtheorem{thm}{Theorem}[section]
\newtheorem{lm}{Lemma}[section]
\newtheorem{ex}{Example}[section]
\newproof{pf}{Proof}
\newtheorem{openpb}{Open Problem}
\newtheorem{cor}{Corollary}[section]

\begin{document}

\begin{frontmatter}
  \title{Bulking II: Classifications of Cellular Automata\tnoteref{phd}}
  \tnotetext[phd]{The results presented here first appeared to a great extent in French in the PhD theses of Ollinger~\cite{Ollinger:2002:PhD} and Theyssier~\cite{Theyssier:2005:PhD}}
  \author[lip]{M. Delorme}
  \author[lip]{J. Mazoyer}
  \author[lif]{N. Ollinger}
  \author[lama]{G. Theyssier\corref{cora}} 
  \cortext[cora]{Corresponding author
    (\url{Guillaume.Theyssier@univ-savoie.fr})}

  \address[lip]{LIP, ENS Lyon, CNRS, 46 all\'ee d'Italie,
    69\hspace{0.2em}007 Lyon, France} \address[lif]{LIF, Aix-Marseille
    Universit\'e, CNRS, 39 rue Joliot-Curie, 13\hspace{0.2em}013
    Marseille, \rlap{France}} \address[lama]{LAMA, Universit\'e de
    Savoie, CNRS, 73\hspace{0.2em}376 Le Bourget-du-Lac Cedex, France}

\begin{abstract}
  This paper is the second part of a series of two papers dealing with
  bulking: a way to define quasi-order on cellular automata by
  comparing space-time diagrams up to rescaling.  In the present
  paper, we introduce three notions of simulation between cellular
  automata and study the quasi-order structures induced by these
  simulation relations on the whole set of cellular automata. Various
  aspects of these quasi-orders are considered (induced equivalence
  relations, maximum elements, induced orders, etc) providing several
  formal tools allowing to classify cellular automata.
\end{abstract}

\end{frontmatter}

\section{Introduction}
\label{sec:intro}

In the first paper \cite{bulking1}, we have developped a general theory of
bulking aimed at defining quasi-orders on cellular automata based on the idea of
space-time rescaling. The present paper focuses on three instances of such
quasi-orders and uses them as classification tools over the set of
one-dimensional cellular automata.

Classifying does not make sense without additional assumptions (some
criteria of classification). If in Wolframs
papers~\cite{Wolfram:1984:CTCA} these criteria were implicit and
informal, several classifications with explicit and formal criteria
have been since
proposed~\cite{Gilman:1987:CLA,Cattaneo:1999:TDD,Kurka97}. Usually,
the criteria are those of dynamical systems and consist in a finite
list of qualitative behaviors. Our approach here is different: we
don't define any \textit{a priori} list of behaviors. Instead, we
consider a simulation relation (a quasi-order) which tells when some
cellular automaton is able to reproduce the behavior of another. The
criterion of classification is then the definition of the
quasi-order. Our central thesis is that, when it comes to apprehend
the great variety of behaviors in cellular automata, the language of
orders (equivalence classes, chains, ideals, maximal elements,
distance to the bottom, etc) is more adapted than a finite list of
monadic predicates (of the form ``having property P'' for some P).

In this paper, we introduce three quasi-orders. They are all defined
according to the same scheme developped in the companion paper
\cite{bulking1}: some local comparison relation up to spatio-temporal
rescaling. They only differ in the local comparison they use, which
are based on the two following basic notions:
\begin{itemize}
\item the \emph{injection} of a small system ($A$) \emph{into} a larger one ($B$),
\item the \emph{projection} of a large system ($B$) \emph{onto} a smaller one ($A$).
\end{itemize}

The three quasi-orders can be defined informally as follows:
\begin{itemize}
\item an \emph{injective} simulation of $A$ by $B$, denoted by $A
  \sacsimu B$, is an injection of some rescaling of $A$ into some
  rescaling of $B$;
\item a \emph{surjective} simulation of $A$ by $B$, denoted by $A
  \facsimu B$, is the projection of some rescaling of $B$ onto some
  rescaling of $A$;
\item a \emph{mixed} simulation of $A$ by $B$, denoted by $A
  \facsacsimu B$, is the injection into some rescaling of $B$ of some
  $C$ that projects onto some rescaling of $A$.
\end{itemize}

In the context of cellular automata, the two notions of local
comparison above (injection and projection) translate into the
following.  The first notion is the \emph{subautomaton} relation ($B$
obtained from $A$ by forgetting some states) and the second one is the
\emph{quotient} relation ($B$ obtained from $A$ by identifying some
sates). 

The subautomaton relation was already introduced in \cite{bulking1}
and its importance in cellular automata is illustrated by theorems 4
and 5 of that paper. 

The quotient relation can be seen as a particular case of the notion
of factor in dynamical systems theory and symbolic dynamics
(homomorphism between shift-commuting continuous global maps,
see~\cite{LindMarcus}). More intuitively, the quotient relation is a
means to extract coarse-grained information ($A$) from a complex
system ($B$) (see \cite{physcoarsegrained}). For instance, the
metaphor of particles moving in a stable background used in the
literature of cellular automata\cite{Boccara} follows this idea: some
information (e.g. the phase of the background) is hidden by
identification of states. However, our definition of quotient requires
that both $B$ (the original system) and $A$ (the one obtained after
identification of some states) are cellular
automata. 


We study these orders with several points of view and aim at
understanding their structure as well as showing that they suitably
capture many classical properties or phenomena of cellular automata.

For instance, concerning the phenomenon of universality, we show that
orders $\sacsimu$ and $\facsacsimu$ have a maximum, that classes
of CA having Turing-universality can be obtained by simulating (in a
way closed to Smith~III~\cite{Smith:1971:SCU}) a universal Turing
machine, and that such a class is not necessarily at the top of the
order.

As another example, we show that many global properties of cellular
automata as dynamical systems (reversibility, sensitivity,
expansivity, etc) or cellular automata as computational devices
(ability to simulate a Turing head, or to propagate some signal) characterize
an ideal or a filter in our orders.

\paragraph{Overview of the paper} Section~\ref{sec:def} introduces
three different comparison relations which are three different
instances of the bulking theory developed in the companion paper
\cite{bulking1}.  Section~\ref{sec:co} sets the definitions of these
three notions of simulations and establishes some of their basic
properties. Section~\ref{sec:bot} studies the 'bottom' of each of the
three quasi-orders induced on CA, \textit{i.e.} CA or classes of CA of
least complexity. Section~\ref{sec:sp} focuses on the order structure
with respect to various classical properties of CA, and from a
computability point of view. Then section~\ref{sec:top} explores the
set of CA at the 'top' of these quasi-orders: universal CA. Once
again, the point of view is both structural and computational.
Finally, section~\ref{sec:io} is devoted to the construction of
noticeable induced orders (like infinite chains), and the study of how
simple families of CA spread over these quasi-orders.

\subsection{Definitions}
\label{sec:def}

In this paper, we adopt the setting of one-dimensional cellular
automata with a canonical neighborhood (connected and centered).  A
cellular automaton (CA) is a triple ${\ACA=(S, r, f)}$ where:
\begin{itemize}
\item $S$ is the (finite) \emph{state set},
\item $r$ is the neighborhood \emph{radius},
\item $f:S^{2r+1}\rightarrow S$ is the \emph{local transition function}.
\end{itemize}

A coloring of the lattice $\ZZ$ with states from $S$
(\textit{i.e.}  an element of $S^\ZZ$) is called a
\emph{configuration}.  To $\ACA$ we associate a global function
$G$ acting on configurations by synchronous and uniform
application of the local transition function.  Formally, $G:
S^\ZZ\rightarrow S^\ZZ$ is defined by:
\[G(x)_z = f(x_{z-r},\ldots,x_{z+r})\] for all ${z\in\ZZ}$.  Several
CA can share the same global function although there are syntactically
different (different radii and local functions). However we are mainly
interested in global functions and will sometimes define CA through
their global function without specifying particular syntactical
representations. In addition, the Curtis-Heldund-Lyndon theorem
\cite{Hedlund:1969} allows us to freely compose global CA functions to
construct new CA without manipulating explicitly the underlying
syntactical representation.

When dealing with several CA simultaneously, we use index notation to
denote their respective state sets, radii and local functions. For
instance, to $\ACA$ we associate $\alphA$, $\radA$ and $\locA$.

This paper will make an intensive use of $\tPmCS$ transforms defined
in section 4.2 of \cite{bulking1}, but restricted to dimension 1. With
this restriction, a $\tPmCS$ transform $\alpha$ has the form
$\alpha=\left\langle m,\tau,T,s\right\rangle$ where $m$ and $T$ are
positive integers, $s$ is a (possibly negative) integer and $\tau$ is
either $1$ or $-1$.

For any CA $\ACA$, we denote by $\bulk{\ACA}{\alpha}$ or more
explicitly $\bulk{\ACA}{m,\tau,T,s}$ the application of $\alpha$ to
$\ACA$, which is, according to notations of \cite{bulking1}, a CA of
state set $\alphA^m$ and global rule:
\[\left\langle\boxplus_m,V_\tau\odot\square_m\right\rangle\circ\sigma_s\circ
G_{\mathcal{A}}^T\circ\left\langle\boxplus_m,V_\tau\odot\square_m\right\rangle^{-1}.\]
To simplify notation we will use a shortcut for purely temporal
transforms: for any CA $\ACA$ we denote by $\ACA^t$ the CA
$\bulk{\ACA}{1,1,t,0}$. Finally, as another special case, we denote
by $\grp{\ACA}{n}$ the grouped instance of $\ACA$ of parameter $n$: it
corresponds to the transform $\left\langle n, 1, n, 0\right\rangle$
(see~\cite{bulking1} for a detailed exposition of grouping).

\section{Canonical orders}
\label{sec:co}

In this section we introduce the three bulking quasi-orders that are
studied all along the paper. They are obtained by applying the bulking
axiomatics developed in the companion paper \cite{bulking1} to three
'canonical' relations between local rules of CA.

Those thee 'canonical' relations are in turn based on two classical
notions of morphism between local transition rules of CA:
sub-automaton and quotient-automaton. As shown below, the three
relations we consider are exactly the reflexive and transitive
relations that can be defined by compositions of one or more such
morphisms.

\subsection{From Three Local Relations to Three Bulking Quasi-Orders}
\label{sec:canondef}

A \emph{sub-automaton} is a restriction of a CA to a stable
sub-alphabet. A \emph{quotient} is a projection of a CA onto a smaller
alphabet and compatible with the local transition rule\footnote{A
  quotient is a particular kind of \emph{factor}, a classical notion
  in dynamical systems theory and symbolic dynamics \cite{kurkabook}}.
Both define a kind of morphism between cellular automata:

\begin{itemize}
\item $\ACA$ is a \emph{sub-automaton} of $\ACB$, denoted
  ${\ACA\sac\ACB}$, if there is an injective map
  ${\iota:\alphA\rightarrow\alphB}$ such that
  ${\unif{\iota}\circ\globA=\globB\circ \unif{\iota}}$, where
  ${\unif{\iota}:\alphA^\ZZ\rightarrow \alphB^\ZZ}$ denotes the
  uniform extension of $\iota$. We often write $\ACA\sacby{\iota}\ACB$
  to make the map $\iota$ explicit.
    
\item $\ACA$ is a \emph{quotient} of $\ACB$, denoted ${\ACA\fac\ACB}$,
  if there is a surjective (onto) map $s$ from $\alphB$ to $\alphA$
  such that ${\unif{s}\circ\globB=\globA\circ \unif{s}}$, where
  ${\unif{s}:\alphB^\ZZ\rightarrow \alphA^\ZZ}$ denotes the uniform
  extension of $s$. We also write $\ACA\facby{s}\ACB$ to make the map
  $s$ explicit.
\end{itemize}

Relations $\sac$ and $\fac$ are quasi-orders (reflexive and
transitive) and it is straightforward to check that their induced
equivalence relation is the relation of isomorphism between cellular
automata (equality up to state renaming) denoted by $\isom$.

It is also straightforward to check that $\sac$ and $\fac$ are
incomparable (none of them is implied by the other one). It is thus
interesting to consider compositions of them. The composition of two
relations $R_1$ and $R_2$ is the relation $R_1\cdot R_2$ defined by
\[R_1\cdot R_2 = \{(x,y) : \exists z, (x,z)\in R_1\text{ and }(z,y)\in
R_2\}.
\]
We denote by $\locrelset$ the set of relations obtained by (finite)
composition of $\fac$ and $\sac$. Any relation of $\locrelset$ is a
priori interesting, but the following theorem justifies that we
restrict to $\fac$, $\sac$ and the composition $\fac\cdot\sac$ only.  In
the sequel $\fac\cdot\sac$ is denoted by $\facsac$ and, as for $\sac$ and
$\fac$, we use the infix notation (${\ACA\facsac\ACB}$).

\begin{thm}
  \label{thm:locrel}
  \par\noindent
  \begin{enumerate}
  \item any relation $R\in\locrelset$ is included in $\facsac$
    (\textit{i.e., }${(\ACA,\ACB)\in R}$ implies ${\ACA\facsac\ACB}$) ;
  \item the transitive relations of $\locrelset$ are exactly: $\fac$,
    $\sac$ and $\facsac$.
  \end{enumerate}
\end{thm}
\begin{pf}
  We first prove that if ${\ACA\sac\cdot\fac\ACB}$ then
  ${\ACA\facsac\ACB}$, which is sufficient to prove assertion 1 by
  transitivity of $\sac$ and of $\fac$.  So consider $\ACA$, $\ACB$
  and $\ACC$ such that ${\ACA\sac_\iota\ACC}$ and
  ${\ACC\fac_s\ACB}$. Then consider $Q=s^{-1}\circ \iota(\alphA)$. We
  have ${\globB(Q^\ZZ)\subseteq Q^\ZZ}$ because
  \newcommand\cmtinproof[1]{\ \text{(#1)}}
  \begin{align*}
    \unif{s}\circ\globB(Q^\ZZ) &= \globC\circ\unif{s}(Q^\ZZ)\cmtinproof{because $\ACC\fac_s\ACB$}\\ &=
    \globC\circ\unif{\iota}(\alphA^\ZZ)\cmtinproof{by definition of $Q$}\\ &=
    \unif{\iota}\circ\globA(\alphA^\ZZ)\cmtinproof{because ${\ACA\sac_\iota\ACC}$}\\ &\subseteq \unif{\iota}
    (\alphA^\ZZ).
  \end{align*}
The CA $\ACX = (Q, \radB, \locB)$
  is thus well-defined and by definition we have
  ${\ACX\sac\ACB}$. Moreover, we have ${\ACA\facby{\iota^{-1}\circ
      s}\ACX}$ because ${\iota^{-1}\circ s : Q \rightarrow A}$ is
  well-defined and onto, and because
  \[\unif{\iota^{-1}\circ s}\circ\globX = \globA \circ\unif{\iota^{-1}\circ s}\]
  since ${\unif{s}\circ\globB = \globC\circ\unif{s}}$ and
  ${\unif{\iota^{-1}}\circ\globC = \globA\circ\unif{\iota^{-1}}}$ over
  ${\bigl(\iota(A)\bigr)^\ZZ = \unif{s}\bigl(Q^\ZZ\bigr)}$.  Hence
  ${\ACA\facsac\ACB}$ and assertion 1 is proven.

  Given assertion 1 we have ${\locrelset = \{\fac,\sac,\sac\cdot\fac,
    \facsac\}}$. To prove assertion 2, it is thus sufficient to prove
  that ${\sac\cdot\fac}$ is not transitive.  To do this, consider
  ${\alphA=\{0,\ldots,p-1\}}$ with $p$ prime, ${p\geq 5}$, and let
  ${\alpha, a_0, a_1, b_0, b_1}$ be five distinct elements of
  $\alphA$. Then consider $\ACA$, the CA with state set $\alphA$,
  radius $1$ and local rule $\locA$ defined by:
  \[\locA(\ast,x,y) =
  \begin{cases}
    a_{1-i} &\text{if } x\not=\alpha\text{ and }y=a_i,\\
    b_{1-i} &\text{if } x\not=\alpha\text{ and }y=b_i,\\
    y+1\bmod p&\text{else.}
  \end{cases}\] $\locA$ depends only on two variables. Suppose now
  that there is some AC $\ACB$ with at least two states such that
  ${\ACB\fac_\pi\ACA}$. We will show that $\pi$ must be one-to-one.
  Suppose for the sake of contradiction that there are distinct
  elements $x$ and $y$ in $\alphA$ such that ${\pi(x)=\pi(y)}$. Then,
  because ${\locA(\ast,\alpha,z)=z+1\bmod p}$ for any $z$, we have
  ${\pi(x+1 \bmod p)=\pi(y+1 \bmod p)}$ and more generally \[\pi(x+i
    \bmod p)=\pi(y+i \bmod p)\] for all ${i\in\NN}$. So, supposing
  without loss of generality ${y>x}$, let ${k=y-x}$. We deduce from
  above that ${\pi(y)=\pi(y+jk\bmod p)}$ for all ${j\in\NN}$ and, by
  elementary group theory, that $\pi$ is constant equal to $\pi(y)$
  (because $p$ is prime and $k\not=0$).  This is in contradiction with
  the fact that $\pi$ has image $\alphB$ which has at least two
  elements. So $\pi$ is one-to-one and $\ACB$ is isomorphic to
  $\ACA$. Now consider $\ACC$, the identity CA over state set
  ${\alphC=\{0,1\}}$. Since $\ACC$ possesses $2$ quiescent states and
  $\ACA$ has no quiescent state (straightforward from the definition
  of $\locA$ above), we have ${\ACC\cancel{\sac}\ACA}$.  With the
  discussion above, we can conclude that
  ${\ACC\cancel{\sac\cdot\fac}\ACA}$.

  However, we have ${\ACB\facsac\ACA}$ because the states
  ${\{a_0,a_1,b_0,b_1\}}$ induce a sub-automaton $\ACC$ of $\ACA$
  which verifies ${\ACB\fac_s\ACC}$ where ${s :
    \{a_0,a_1,b_0,b_1\}\rightarrow\{0,1\}}$ is defined by ${s(a_i)=0}$
  et ${s(b_i)=1}$. Assertion 2 follows since the relation $\facsac$ is
  included in the composition of the relation $\sac\cdot\fac$ with
  itself. \qed
\end{pf}

Like $\sac$ (already considered in \cite{bulking1}), $\fac$ and
$\facsac$ are quasi-orders on CA and therefore constitute natural
candidates for the $\opdiv$ relation of bulking axiomatics (definition
8 of \cite{bulking1}).

Inspired by definition 14 of \cite{bulking1}, we now define 3 bulking
quasi-orders using $\tPmCS$ transforms.  

\begin{defn}
  $\ACB$ simulates $\ACA$ \emph{injectively}, denoted
  ${\ACA\sacsimu\ACB}$, if there exist two $\tPmCS$ transforms
  $\alpha$ and $\beta$ such that
  ${\bulk{\ACA}{\alpha}\sac\bulk{\ACB}{\beta}}$.
\end{defn}

We will occasionally use the notion of simulation by grouping
introduced in \cite{Mazoyer:1999:IOC} and discussed in
\cite{bulking1}: we denote by ${\ACA\grporder\ACB}$ the fact that
there are $n$ and $m$ such that
${\grp{\ACA}{n}\sac\grp{\ACB}{m}}$. This is a special case of the
injective simulation above.

\begin{defn}
  $\ACB$ simulates $\ACA$ \emph{surjectively}, denoted
  ${\ACA\facsimu\ACB}$, if there exist two $\tPmCS$ transforms
  $\alpha$ and $\beta$ such that
  ${\bulk{\ACA}{\alpha}\fac\bulk{\ACB}{\beta}}$.
\end{defn}

\begin{defn}
  $\ACB$ simulates $\ACA$ in a mixed way, denoted
  ${\ACA\facsacsimu\ACB}$, if there exist two $\tPmCS$ transforms
  $\alpha$ and $\beta$ such that
  ${\bulk{\ACA}{\alpha}\facsac\bulk{\ACB}{\beta}}$
\end{defn}

For each notion of simulation above, we say that the simulation is
\emph{strong} if the transformation $\alpha$ applied to the simulated
CA is trivial: ${\alpha=\langle 1,1,1,0\rangle}$ so that
${\bulk{\ACA}{\alpha}=\ACA}$.

\begin{thm}
  $\bigl(CA,\sacsimu\bigr)$, $\bigl(CA,\facsimu\bigr)$ and
  $\bigl(CA,\facsacsimu\bigr)$ are quasi-orders.
\end{thm}
\begin{pf}
  We show that $\sacsimu$ and $\facsacsimu$ correspond exactly to
  models of bulking developped in \cite{bulking1}: the proof of
  theorem 15 of \cite{bulking1} contains the case of injective
  simulation. The case of $\facsacsimu$ follows immediately (axiom
  $(B_4)$ is straightforward and axiom $(B_5)$ is verified because
  $\facsac$ contains $\sac$). For $\facsimu$, the proof of each axiom
  is similar except for axiom $(B_5)$.

  
  With or without axiom $(B_5)$, theorem 10 of \cite{bulking1} can be
  applied in each case and show the present theorem.\qed
\end{pf}

\begin{lm}
  \label{lem:mapa}
  Let $\triangleleft$ be any relation among $\sac$, $\fac$ and $\facsac$. Then the following
  propositions are equivalent:
  \begin{itemize}
  \item there exist two $\tPmCS$ transforms $\alpha$ and $\beta$ such that
    ${\bulk{\ACA}{\alpha}\,\triangleleft\,\bulk{\ACB}{\beta}}$,
  \item there exist a $\tPmCS$ transform $\alpha$ and an integer $t$ such that
    ${\bulk{\ACA}{\alpha}\,\triangleleft\,\grp{\ACB}{t}}$,
  \item there exist a $\tPmCS$ transform $\beta$ and an integer $t$ such that
    ${\grp{\ACA}{t}\,\triangleleft\,\bulk{\ACB}{\beta}}$,
  \end{itemize}
\end{lm}
\newcommand\trsfo[1]{\left\langle #1, 0\right\rangle}
\begin{pf}
  We use the property of compatibility of relation $\triangleleft$ with respect
  to geometrical transforms (axiom $B_4$ of \cite{bulking1}).  The lemma follows
  from the following property: for any transform $\alpha$, there exist a
  transform $\beta$ and an integer $t$ such that
  \[\forall F: \bulk{\bigl(\bulk{F}{\alpha}\bigr)}{\beta} = \grp{F}{t}.\]
  If $\alpha=\trsfo{m,t,z}$, $\beta$ can be chosen as the composition of
  $\trsfo{1,m,0}$, $\trsfo{1,1,-z}$ and $\trsfo{t,1,0}$.\qed
\end{pf}

In the sequel, if $\simu$ denotes a simulation quasi-order we denote
by $\sime$ the induced equivalence relation and by $\simc{\ACA}$ the
equivalence class of $\ACA$ with respect to $\sime$. For instance, to
$\sacsimu$ we associate the notations $\sacsime$ and
$\sacsimc{\ACA}$ with the following meanings:
\begin{align*}
  \ACA\sacsime\ACB &\iff \ACA\sacsimu\ACB\text{ and }\ACB\sacsimu\ACA,\\
  \sacsimc{\ACA} &= \{\ACB : \ACA\sacsime\ACB\}.
\end{align*}
 We use similar notations for $\facsimu$ and
$\facsacsimu$.

Before entering into details concerning various aspects of the three
simulation relations defined above, we can already make a clear (yet
informal) distinction between $\sacsimu$ and $\facsacsimu$ on one hand,
and $\facsimu$ on the other hand. For the two former, the simulation
takes place on a subset of configurations and nothing can be said
\textit{a priori} about the behavior of the simulator outside this
subset of configurations. For $\facsimu$, however, the simulation
occurs on any configuration and the simulator's behavior on any
configuration is in some way affected by the
simulation. Section~\ref{sec:idfi} give several illustrations of this
difference.

\subsection{First Properties}
\label{sec:firstprop}

We now establish a set of basic general facts about $\sacsimu$,
$\facsimu$ and $\facsacsimu$ while next sections of the paper focus on
particular aspects.

\begin{thm}
  \label{thm:bfacts}
  Let $\ACA$ be any CA and $\simu$ be any relation among $\sacsimu$,
  $\facsimu$ and $\facsacsimu$. Then it holds:
  \begin{enumerate}
  \item there is some $\ACB\in\simc{\ACA}$ having a quiescent state,
  \item there is some $\ACB\in\simc{\ACA}$ with radius $1$,
  \item ${\sing\simu\ACA}$ where $\sing$ is the CA with a single state,
  \item ${\ACA\simu\ACA\times\ACB}$ and ${\ACA\simu\ACB\times\ACA}$ for any $\ACB$.
  \end{enumerate}
\end{thm}
\begin{pf}
  \par\noindent
  \begin{enumerate}
  \item there exists some uniform configuration $x$ and some ${t\geq
      1}$ such that $\globA^t(x)=x$. So $\ACA^t$ has a quiescent
    state and it clearly belongs to $\simc{\ACA}$.
  \item $\bulk{\ACA}{\radA,1,1,0}$ admits a syntactical representation
    with radius 1 and clearly belongs to $\simc{\ACA}$.
  \item First, one always has ${\sing\fac_\pi\ACA}$ where $\pi$ is the
    trivial surjection mapping each state of $\ACA$ to the single
    state of $\sing$. So assertion 3 is proven for $\facsimu$ and
    $\facsacsimu$. Second, one has ${\sing\sac_i\ACB}$ if $\ACB$ has a
    quiescent state where $i$ is the trivial injection mapping the
    single state of $\sing$ to the quiescent state of
    $\ACB$. Assertion 3 follows for $\sacsimu$ by assertion 1.
  \item We show only the first relation, the second being rigorously
    symmetric. First, one has always
    ${\ACA\fac_{\pi_1}\ACA\times\ACB}$ where ${\pi_1 :
      \alphA\times\alphB\rightarrow \alphA}$ is the projection over
    the first component. Second, if $\ACB$ has a quiescent state $q$,
    one has ${\ACA\sac_\iota\ACA\times\ACB}$ where $\iota$ is the
    injection defined by ${\iota(x) = (x,q)}$ for all ${x\in\alphA}$
    (the equality ${\unif{\iota}\circ\globA =
      (\globA\times\globB)\circ\unif{\iota}}$ is true over
    $\alphA^\ZZ$). If $\ACB$ has no quiescent state, just consider
    $\ACB^t$ and apply the previous reasoning to obtain:
    \[\ACA^t\sac\ACA^t\times\ACB^t=\bigl(\ACA\times\ACB\bigr)^t\]
    and thus ${\ACA\sacsimu\ACA\times\ACB}$.
  \end{enumerate}
  \qed
\end{pf}

The three simulation quasi-orders are derived through bulking
axiomatics from three different relations on local rules
(see~\ref{thm:locrel}). There is \textit{a priori} no reason why the
differences between local relations should extend to differences
between the three simulation quasi-orders. The following theorem shows
that $\sacsimu$, $\facsimu$ and $\facsacsimu$ are nevertheless
different and that $\sacsimu$ and $\facsimu$ are both strictly
included in $\facsacsimu$.

\begin{thm}
  The relations $\sacsimu$ and $\facsimu$ are incomparable (no
  inclusion in either direction).
\end{thm}
\begin{pf}
  We first show that there are CA $\ACA$ and $\ACB$ such that
  ${\ACA\sac\ACB}$ but ${\ACA\cancel\facsimu\ACB}$.  Let ${\ACA =
    \sigma\times\sigma^{-1}}$ defined over states set ${\alphA =
    \{0,1\}\times\{0,1\}}$ and let $\ACB$ be the CA of radius $1$
  defined over ${\alphB = \alphA\cup\{\#\}}$ by:
  \[\locB(x,y,z) =
  \begin{cases}
    \locA(x,y,z) &\text{if $x,y,z\in\alphA$,}\\
    y &\text{else.}
  \end{cases}\]
  One clearly has ${\ACA\sac_{Id}\ACB}$. Now suppose
  ${\ACA\facsimu\ACB}$. Without loss of generality we can assume
  that there are geometrical transforms ${\alpha=<m,\tau,T,s>}$ and
  ${\beta=<m',1,T',0>}$ such that
  ${\bulk{\ACA}{\alpha}\fac_\pi\bulk{\ACB}{\beta}}$. But, by
  definition of $\ACB$, there exists some state $q_0$ of
  $\bulk{\ACB}{\beta}$ (for instance $\#^{m'}$) which is left
  invariant by iteration of ${\bulk{\ACB}{\beta}}$ whatever the
  context. Then $\pi(q_0)$ must be a state of $\bulk{\ACA}{\alpha}$
  with the same property. This is impossible since either
  ${s\not=T}$ or ${s\not=-T}$ and thus some component of the future
  state of a cell of $\bulk{\ACA}{\alpha}$ is dependent of the state
  of a neighboring cell.

  Now we show that there are $\ACA$ and $\ACB$ such that
  ${\ACA\fac\ACB}$ but ${\ACA\cancel\sacsimu\ACB}$ and the theorem
  follows.  Let $\ACA$ and $\ACB$ be the automata pictured on
  figure~\ref{fig:diagab}. $\ACA$ is a CA with two states, $0$ and
  $1$, whose behavior is to reduce ranges of $1$'s progressively until
  they reach size $1$: at each time step the cells at each ends of a
  range of size $3$ or more are turned into state $0$ (only the right
  cell of range of size $2$ is turned into $0$). $\ACB$ has three
  states ($0$, $1$ and $2$) and has the following behavior: ranges of
  size $3$ or more of non-zero states are reduced in a similar way by
  the two ends (states inside ranges are left unchanged), ranges of
  size $2$ become an isolated $2$ (left cell becomes $2$ and right
  cell $0$), and ranges of size $1$ become an isolated $1$. In a word,
  $\ACB$ reduces the size of non-zero ranges until size $1$ but keeps
  the parity information at the end: an even range becomes eventually
  an isolated $2$ and an odd range becomes an isolated $1$ (see
  figure~\ref{fig:diagab}).

  Formally, let ${\pi : \{0,1,2\}\rightarrow \{0,1\}}$ be the
  surjective function defined by ${\pi(0)=0}$ and ${\pi(x)=1}$ if
  ${x\not=0}$. Now let $\ACA$ be the CA of radius $2$ and state set
  ${\alphA=\{0,1,2\}}$ with local rule:
  \[\locA(x,y,z,t,u)=
  \begin{cases}
    1&\text{if }\overline\pi(xyztu)=01110,\\
    2&\text{if }\overline\pi(yztu)=0110,\\
    z&\text{if }\overline\pi(yzt)=111\text{ and if }\overline\pi(xyztu)\not=01110,\\
    z&\text{if }\overline\pi(yzt)=010,\\
    0&\text{in any other case.}
  \end{cases}\] Finally, let $\ACB$ be the CA with states set
  ${B=\{0,1\}}$, radius $2$ and local transition function
  \[\locB(x,y,z,t,u)=
  \begin{cases}
    1&\text{if }yzt=111\text{ or }yzt=010\text{ or }yztu=0110,\\
    0&\text{else.}
  \end{cases}\] 
  \begin{figure}[htbp]
    \centering
    \begin{tabular}{cc}
      \includegraphics[width=.4\textwidth]{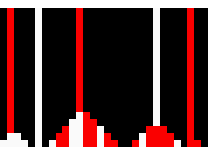} & \includegraphics[width=.4\textwidth]{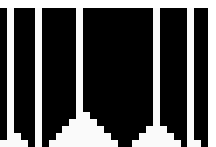}
    \end{tabular}
    \caption{\label{fig:diagab}Behavior of $\ACA$ (left) and $\ACB$
      (right). Time goes from bottom to top.}
  \end{figure}
  By construction, we have ${\ACA\fac_\pi\ACB}$. Now suppose for the
  sake of contradiction that ${\ACA\sacsimu\ACB}$ and more precisely:
  \[\bulk{\ACA}{m',\tau',t',s'}\sac_\phi\bulk{\ACB}{m,1,t,0},\]
  where ${\alpha =<m',\tau',t',s'>}$ and ${\beta=<m,1,t,0>}$ are
  suitable geometrical transforms.  Let ${u=1^{m'}}$ and ${v=0^{m'}}$
  ($u$ and $v$ are particular states of $\bulk{\ACA}{\alpha}$) and
  consider ${U=\phi(u)}$ and ${V=\phi(v)}$ ($U$ and $V$ belong to
  ${\alphB^{m'}}$). The remaining of the proof below proceeds by a
  careful case analysis on $U$ and $V$ to obtain a final
  contradiction. The main technique is to consider specific orbits of
  $\bulk{\ACA}{\alpha}$ involving $u$ and $v$, and to derive
  constraints on their possible image by $\phi$ involving $U$ and $V$.

  Since configurations $\unif{u}$ and $\unif{v}$ are fixed points of
  $\bulk{\ACA}{\alpha}$, so are $\unif{U}$ and $\unif{V}$ for
  ${\bulk{\ACB}{\beta}}$. Moreover, one can check from the definition
  above that the state $0$ is a 'blocking state' for $\ACB$: the
  half-configuration on the left of an occurrence of $0$ evolves
  independently of the half-configuration on its right. So, if $U$
  contains one or more zero's, then any configuration of
  $\bulk{\ACB}{\beta}$ containing $U^3$ will contain at least one
  occurrence of $U$ for ever (because it is the case for the
  configuration $\unif{U}$): this is in contradiction with the fact
  that the orbit of a configuration of the form ${{}^\omega v u^4
    v^\omega}$ does not contain any occurrence of $u$ after sufficiently
  many iterations of $\bulk{\ACA}{\alpha}$ (because, whatever the
  value of $m'$, $u^4$ represents in $\ACA$ an even-sized range of 1s
  which is reduced until the last two 1s are turned into a single 2 by
  case 2 of the definition of $\ACA$). Hence we have
  ${\unif{\pi}(U)=1^{m}}$.

  From this we deduce that ${V=0^m}$ because configurations of the
  form ${{}^\omega V U^n V^\omega}$ are transformed into
  configurations where a single cell is not in state $V$ (just
  consider the orbit of ${{}^\omega v u v^\omega}$ under
  $\bulk{\ACA}{\alpha}$) and large ranges of non-zero states are
  always turned into large ranges of zero's under $\ACB$.

  Finally, we have ${\unif{\pi}\bigl(\phi(0^{m'-1}1)\bigr) =
    0^{m-1}1}$ by considering the orbit of a configuration of the
  form ${{}^\omega v u^2 v^\omega}$ under $\bulk{\ACA}{\alpha}$ and
  its counterpart of the form ${{}^\omega V U^2 V^\omega}$ under
  $\bulk{\ACB}{\beta}$ (by the way, we also show that the shift
  parameter of transform $\alpha$ is $0$). Now, letting
  ${u'=0^{m'-1}1}$, we have on one hand the orbits of $2$
  configurations of the form ${{}^\omega v u' u^{2n} v^\omega}$ and
  ${{}^\omega v u^{2n} v^\omega}$ both leading to the same
  configuration of the form ${{}^\omega v u' v^\omega}$ under
  $\bulk{\ACA}{\alpha}$, and on the other hand, the orbits of
  ${{}^\omega V \phi(u') U^{2n}V^\omega}$ and ${{}^\omega V
    U^{2n}V^\omega}$ leading to different fixed points under
  $\bulk{\ACB}{\beta}$ due to different parity of non-zero ranges:
  this is a contradiction since
  ${\unif{\phi}\circ\bulk{\ACA}{\alpha} =
    \bulk{\ACB}{\beta}\circ\unif{\phi}.}$
\qed
\end{pf}

\section{Bottoms of the Orders}
\label{sec:bot}

This section focuses on the bottom of the orders.  We have already
seen (theorem~\ref{thm:bfacts}) that $\sing$ is a global minimum for
the three quasi-orders considered here. In this section, we study CA
that are at the lowest \emph{levels} of the quasi-orders.  Formally,
the only CA at level $0$ is $\sing$ and a CA $\ACA$ is at level
${n+1}$ for a quasi-order $\simu$ if:
\begin{enumerate}
\item $\ACA$ is not at level $n$ and,
\item ${\forall\ACB: \ACB\simu\ACA\Rightarrow \ACB\in\simc{\ACA}\text{
    or $\ACB$ is at level $i$ with ${i\leq n}$.}}$
\end{enumerate}

The following theorem shows that some classical properties of CA
correspond to classes at level 1.  Recall that a cellular automaton is
nilpotent if all initial configurations lead to the same configuration
after a finite time.

\begin{thm}
  \label{thm:nilper}
  Let $\simu$ be a simulation relation among $\sacsimu$, $\facsimu$
  and $\facsacsimu$. Then the following CA are at level 1 (provided
  they have $2$ or more states):
  \begin{enumerate}
  \item the set of nilpotent CA, which is an
    equivalence class for $\sime$,
  \item the set of CA which are periodic up to translation
    (${\ACA^t\circ\sigma_z = Id}$) which is exactly the equivalence
    class for $\sime$ of the identity CA.
  \end{enumerate}
\end{thm}
\begin{pf}
  \par\noindent
  \begin{enumerate}
  \item Nilpotency is equivalent to the existence of a uniform
    configuration reached in a fixed finite time from any
    configuration. This property of phase space is clearly invariant
    by geometrical transforms and preserved by taking sub-automata or
    quotient automata. So any nilpotent CA is at level at most
    $1$. Moreover, the set of nilpotent CA forms an equivalence
    class. Indeed, for any nilpotent $\ACA$, there is $t$ such that
    $\ACA^t$ is a constant function equal to some $\unif{q_a}$.  If we
    consider any nilpotent $\ACB$ with at least $2$ states, there is
    $m$ such that ${\card{\alphB^m}\geq\card{\alphA}}$ and $t'$ such
    that $\ACB^{t'}$ is a constant function equal to some
    $\unif{q_b}$. If we consider the geometrical transforms
    ${\alpha=<1,1,t,0>}$ and ${\beta=<m,1,t',0>}$, then we have both
    ${\bulk{\ACA}{\alpha}\sac_i\bulk{\ACB}{\beta}}$ and
    ${\bulk{\ACA}{\alpha}\fac_\pi\bulk{\ACB}{\beta}}$ if $i$ is such
    that ${i(q_a)=q_b}$ and $\pi$ is such that ${\pi(x)=q_a\iff
      x=q_b}$.
  \item Any CA which is periodic up to a translation is by definition
    equivalent to some identity CA and two identity CA with different
    state set are also clearly equivalent. Moreover, all such CA are
    at level 1 because the property of being periodic up to
    translations is preserved by geometrical transformations and by
    taking sub-automata or quotient automata. \qed
  \end{enumerate}
\end{pf}

In the remaining part of this section, we will study two families of
cellular automata with respect to the quasi-orders: a subset of
additive CA and products of shifts. Our goal is to show that at
(almost) each finite level there are infinitely many incomparable
classes (theorem~\ref{thm:zpz} and corollary~\ref{cor:infinishi} below).

\subsection{Additive Cellular Automata}
\label{sec:botadd}

The bottom of the quasi-order $\grporder$ was studied in
\cite{Mazoyer:1999:IOC}. The main result is the existence of an infinite familly
of mutually incomparable CA at level 1: the familly of CA $\zee{p}$
with $p$ a prime number and where $\zee{p}$ is a CA of radius $1$ and
state set ${\{0,\ldots,p-1\}}$ defined by the following local rule:
\[\delta_{\zee{p}}(x,y,z) = x+y+z \bmod p.\]

There are strong connections between $\grporder$ and $\sacsimu$ and in
fact the set of CA at level $1$ are the same for these two
quasi-orders.

\begin{lm}
  \label{lm:levelone}
  If $\ACA$ is at level $1$ for $\grporder$ then $\ACA$ is at level
  $1$ for $\sacsimu$.
\end{lm}
\begin{pf}
  If ${\ACB\sacsimu\ACA}$ then by lemma~\ref{lem:mapa} there is
  some integer $t$ and some transform $\beta$ such that
  ${\bulk{\ACB}{\beta}\sac\grp{\ACA}{t}}$. By theorem~\ref{thm:bfacts}
  we can suppose that $\ACA$ has radius $1$ so $\bulk{\ACB}{\beta}$
  has radius 1. Since $\ACA$ is at level $1$ for $\grporder$, then
  either ${\bulk{\ACB}{\beta}\in\grpclass{\ACA}}$ or
  ${\bulk{\ACB}{\beta}\in\grpclass{\sing}}$. We deduce that either
  ${\ACB\in\sacsimc{\ACA}}$ or ${\ACB\in\sacsimc{\sing}}$. Hence
  $\ACA$ is at level at most $1$ for $\sacsimu$ and it cannot be at
  level $0$ since it is not in ${\sacsimc{\sing}=\grpclass{\sing}}$.
  \qed
\end{pf}

The previous lemma is not enough to show that the CA
${\bigl(\zee{p}\bigr)_p}$ with $p$ prime are mutually
$\sacsimu$-incomparable because several equivalence classes for
$\grporder$ can be included in a single class for $\sacsimu$.  However
we are going to show that this familly is a set of mutually
incomparable CA for the three quasi-orders introduced
above\footnote{The proof we give here was suggested by
  E.~Jeandel.}. Moreover, for $\grporder$ and $\sacsimu$, they are all
at level $1$. The proof relies on the following result already used
for the case of $\grporder$.

A CA ${(S,r,f)}$ is \emph{LR-permutative} if the two following
functions are bijections for all ${a_1,\ldots,a_{2r}}$:
\begin{itemize}
\item ${x\mapsto f(a_1,\ldots,a_{2r},x)}$ and
\item ${x\mapsto f(x,a_1,\ldots,a_{2r})}$.
\end{itemize}

\begin{thm}[\cite{Mazoyer:1998:ACA}]
  \label{thm:mazrapadd}
  Let $p$ be a prime number and ${t\geq 1}$. Then we have:
  \begin{enumerate}
  \item $\grp{\zee{p}}{t}$ is LR-permutative;
  \item if ${\ACA\sac\grp{\zee{p}}{t}}$ then $p$ divides $\card{\alphA}$.
  \end{enumerate}
\end{thm}

To take into account use of $\fac$ in simulation we will use the
following lemma.

\begin{lm}
  \label{lm:manulem}
  If $\ACB$ is LR-permutative and ${\ACA\fac\ACB}$ then
  $\card{\alphA}$ divides $\card{\alphB}$.
\end{lm}
\begin{pf}
  To simplify notations, we suppose that $\ACB$ is of radius $1$ (the
  proof works the same way for higher radii).  Suppose
  ${\ACA\fac_\pi\ACB}$. By surjectivity of $\pi$, it is sufficient to
  show that $\pi$ is balanced, \textit{i.e.}  such that for all
  ${x,y\in\alphA}$:
  \[\card{\{e : \pi(e) = x\}} = \card{\{e : \pi(e) = y\}}.\]
  Consider any ${x,y\in\alphA}$. Let $a,b\in\alphB$ be such that
  ${\pi(a)=x}$ and ${\pi(b)=y}$ and consider any $c\in\alphB$. By
  R-permutativity there is $d\in\alphB$ such that ${\locB(a,c,d) =
    b}$. Now for any ${a'\in\alphB}$ such that ${\pi(a')=\pi(a)}$, we
  must have ${\pi\bigl(\locB(a',c,d)\bigr)=\pi(b)}$ because
  ${\ACA\fac_\pi\ACB}$. Moreover, by L-permutativity,
  ${a'\mapsto\locB(a',c,d)}$ is one-to-one which proves:
  \[\card{\{a : \pi(a) = x\}} \leq \card{\{b : \pi(b) = y\}}.\]
  The balance of $\pi$ follows by symmetry.  \qed
\end{pf}

The results above are the key ingredient of the following theorem.

\begin{thm}
  \label{thm:zpz}
  Let $\simu$ be any relation among $\sacsimu$, $\facsimu$ and
  $\facsacsimu$.  Let $p$ and $q$ be two distinct prime numbers. Then
  we have:
  \begin{enumerate}
  \item $\zee{p}\not\simu\zee{q}$
  \item $\zee{p}$ is at level $1$ for $\sacsimu$;
  \end{enumerate}
\end{thm}
\begin{pf}
  $2$ follows immediately from lemma~\ref{lm:levelone} and the fact
  that $\zee{p}$ is at level 1 for $\grporder$ (corollary~2 of
  \cite{Mazoyer:1998:ACA}).  To prove assertion $1$ it is enough to
  prove ${\zee{p}\cancel\facsacsimu\zee{q}}$.  Suppose for the sake of
  contradiction that ${\zee{p}\facsacsimu\zee{q}}$, or equivalently by
  lemma~\ref{lem:mapa}, that there are a CA $\ACA$, a transform
  $\alpha$ and an integer $t$ such that
  ${\bulk{\zee{p}}{\alpha}\fac\ACA\sac\grp{\zee{q}}{t}}$. Then,
  combining lemma~\ref{lm:manulem} and theorem~\ref{thm:mazrapadd}, we
  deduce that the number of states of $\bulk{\zee{p}}{\alpha}$ is a
  power of $q$ which contradicts the fact that $p$ and $q$ are two
  distinct primes.  \qed
\end{pf}

\subsection{Products of Shifts}
\label{sec:botshi}

We will now study products of shifts in order to show that there are
infinitely many incomparable CA at any finite level greater than 3 for
any of the three simulation quasi-orders of the paper.

We denote by ${\shi{n}{z}}$ the translation CA with $n$ states
${\{1,\ldots,n\}}$ and translation vector $z$ defined by:
\[\shi{n}{z}(c)_{z'} = c_{z'-z}.\]
We then consider cartesian products of such CA. Since
${\shi{n}{z}\times\shi{p}{z}\isom\shi{np}{z}}$, we can focus on
considering cartesian products where all vectors are distinct.

The next lemma shows that the structure of product of translations is
preserved when taking sub-automata and quotient-automata.

\begin{lm}
  \label{lm:soushi}
  Let $\ACB = \prod_{i=1}^p \shi{n_i}{z_i}$ (with $z_i$ all distinct)
  and suppose $\ACA$ is such that $\ACA\facsac\ACB$ then $\ACA$ is
  isomorphic to ${\prod_{j=1}^k\shi{n'_{i_j}}{z_{i_j}}}$ where ${1\leq
    i_j\leq n}$ and ${2\leq n'_{i_j}\leq n_{i_j}}$.
\end{lm}
\begin{pf}
  The lemma is straightforward if we replace $\facsac$ by $\sac$. So
  it is enough to show that it is also true when replacing $\facsac$
  by $\fac$. Suppose that ${\ACA\fac_\pi\ACB}$.  The idea of the proof
  is to show that $\pi$ must be 'compatible' with the product
  structure: it forgets some components and keeps others but never
  introduces any kind of 'correlation' between them.  So let $i$ be
  such a component in $\ACB$ (${1\leq i\leq p}$). Consider any pair of
  states ${q,q'\in\alphB}$ such that ${q_i = q'_i}$ (where $q_i$ or
  $q'_i$ denotes the projection on the $i$th component). Denote by
  $q_+$ and $q'_+$ the states obtained from $q$ and $q'$ by changing
  their $i$th component in the same way ($(q_+)_i=(q'_+)_i$). 
  Since the $z_i$ are distinct, one can build two
  configurations $c$ and $c'$ of $\ACB$ such that:
  \begin{itemize}
  \item $c(z)=c'(z)$ for all $z\not=0$,
  \item $c(0)=q$ and $c'(0)=q'$,
  \item $\ACB(c)_0=q_+$ and $\ACB(c')_0=q'_+$.
  \end{itemize}
  If $\pi(q)=\pi(q')$ we have $\overline\pi(c)=\overline\pi(c')$ so
  $\pi(q_+) = \pi(q'_+)$. The same reasoning can be done starting from
  $q_+$ and $q'_+$ so we have:
  \[\pi(q) = \pi(q') \iff \pi(q_+) = \pi(q'_+)\]
  
  Hence, if there exist two states $e$ and $e'$ with the same image by
  $\pi$ and which agree on all components except component $i$, then
  values $e_i$ and $e'_i$ can be exchanged in the $i$th component of
  any state without affecting its image by $\pi$.  In such a case, we
  can consider the CA $\ACC$ obtained from $\ACB$ by identifying $e_i$
  and $e_i'$ in the $i$th component. More precisely, $\ACC$ is of the form 
  \[\ACC = \prod_{j=1}^p \shi{n'_j}{z_j}\]
  with $n'_i = n_i - 1$ and $n'_j=n_j$ for any $j\not=i$.  Then we
  have ${\ACA\fac\ACC\fac\ACB}$.  Applying this reasoning iteratively,
  we finally have ${\ACA\fac_g\ACC_0\fac\ACB}$ where $\ACC_0$ is of
  the form ${\prod_{j=1}^k\shi{n'_{i_j}}{z_{i_j}}}$ where ${1\leq
    i_j\leq n}$ and ${2\leq n'_{i_j}\leq n_{i_j}}$ (component reduced
  to $1$ state during one step of the process can be eliminated) and
  $g$ is such that changing the value of any component of any state of
  $\ACC_0$ will change its image by $g$. Now suppose for the sake of
  contradiction that $g$ is not injective. Then there are states $q$
  and $q'$ of $\ACC_0$ such that ${g(q)=g(q')}$. Let ${c=\unif{q}}$
  and $c'$ be equal to $c$ except on position $0$ where it is in state
  $q'$. By hypothesis, at any position $z$, $\glob{\ACC_0}(c)$ and
  $\glob{\ACC_0}(c')$ must be in states having the same image by
  $g$. But since $\ACC_0$ is a product of translations with distinct
  vectors, there must be some position $z$ where $\glob{\ACC_0}(c)$
  and $\glob{\ACC_0}(c')$ are in states which differ on one component
  only: this is in contradiction with the hypothesis on $g$. Hence $g$
  is injective and therefore ${\ACA\isom\ACC_0}$.  \qed
\end{pf}

We will now study the effects of geometrical transformations on CA
which are products of shifts. Of course, the translation vectors
involved in such an automaton can be altered by geometrical
transformations.  If $\ACA$ is a product of translations with vectors
${z_1<\ldots< z_a}$, we denote by $\carashi{\ACA}$ the following
characteristic sequence (provided ${a\geq 3}$):
\[\carashi{\ACA} = \left(\frac{z_3 - z_1}{z_2-z_1}, \ldots, \frac{z_a
    - z_1}{z_2 - z_1}\right).\] The purpose of the following theorem
and lemma is to establish that the characteristic sequence gives a
simple way to compare any pair of products of shifts in the three
quasi-orders.

\begin{thm}
  \label{thm:shishi}
  Let $\simu$ be a relation among $\sacsimu$, $\facsimu$ and
  $\facsacsimu$.  Let $\ACA$ be a product of ${a\geq 3}$ translations
  with distinct vectors and with characterisitic sequence
  ${\carashi{\ACA}=(\alpha_1,\ldots,\alpha_{a-2})}$.  If
  ${\ACB\simu\ACA}$ then $\ACB$ is equivalent to some $\ACC$ which is
  a product of a subset of $b$ translations of $\ACA$.  Moreover, we
  have the following properties: 
  \begin{enumerate}
  \item if ${b=a}$ then $\ACC$ has the same characteristic sequence
    than $\ACA$;
  \item if ${b=a-1}$ and ${b\geq 3}$ then the characteristic sequence
    of $\ACC$ has one of the following form:
    \begin{itemize}
    \item ${(\alpha_1,\ldots,\alpha_{i-1},\alpha_{i+1},\ldots,\alpha_{a-2})}$
    \item
      ${\left(\frac{\alpha_2}{\alpha_1},\ldots,\frac{\alpha_{a-2}}{\alpha_1}\right)}$
    \item
      ${\left(\frac{\alpha_2-1}{\alpha_1-1},\ldots,\frac{\alpha_{a-2}-1}{\alpha_1-1}\right)}$
    \end{itemize}
  \item if the characteristic sequence of $\ACC$ is not $\carashi{\ACA}$ then
    ${\ACA\cancel\simu\ACC}$.
  \end{enumerate}

\end{thm}
\begin{pf}
  Let ${z_1<z_2<\ldots< z_a}$ be the ordered list of translation
  vectors of $\ACA$. Since ${\ACB\simu\ACA}$, there is some $\ACC$
  equivalent to $\ACB$ and some integer ${t\geq1}$ such that
  ${\ACC\facsac\grp{\ACA}{t}}$ (by lemma~\ref{lem:mapa}).  We
  deduce from lemma~\ref{lm:soushi} that $\ACC$ is isomorphic to a
  product of translations whose vectors are a subset of the familly
  $(z_i)$ since $\ACA$ and $\grp{\ACA}{t}$ have identical translation
  vectors, $\ACC$ must have the same characteristic sequence than
  $\ACA$ if it has the same number of translation vectors.  When
  ${b=a-1}$ and ${b\geq 3}$, it is straightforward to check that the
  three possible forms of the characteristic sequence of $\ACC$
  correspond to the case where the missing vector is $z_i$, $z_2$ and
  $z_1$ respectively. To prove the last assertion of the theorem, it
  is sufficient to check that for any transform $\alpha$ of the form
  $<m,1,mt,mz>$ (we can restrict to such transforms by lemma~\ref{lem:mapa}, $\ACC$ and $\bulk{\ACC}{\alpha}$ are
  products of translations with the same characteristic sequence
  because each vector $z_i$ of $\ACC$ becomes ${tz_i+z}$ in
  $\bulk{\ACC}{\alpha}$. \qed
\end{pf}

The next lemma gives canonical members of the equivalence classes of
products of shifts.

\begin{lm}
  \label{lm:canoshi}
  Let $\sime$ be the equivalence relation induced by any of the
  quasi-order $\sacsimu$, $\facsimu$ and $\facsacsimu$.  Consider any
  ${t\not=0}$, any $z$ and any product of translations of the form
  $\ACA=\prod_{i=1}^p \shi{n_i}{tz_i+z}$, . Then we have:
  \[\ACA\sime\prod_{1\leq i\leq p} \shi{2}{z_i}\]
\end{lm}
\begin{pf}
  Let ${\ACB=\prod_{i=1}^p \shi{2}{z_i}}$ and let ${m=\max n_i}$. It is
  straightforward to check that ${\ACA\fac\bulk{\ACB}{m,1,mt,mz}}$ and
  ${\ACA\sac\bulk{\ACB}{m,1,mt,mz}}$, and also that
  ${\bulk{\ACB}{1,1,t,z}\fac\ACA}$ and ${\bulk{\ACB}{1,1,t,z}\sac\ACA}$.  \qed
\end{pf}

Theorem~\ref{thm:shishi} and lemma~\ref{lm:canoshi} give a complete
characterisation of the position of products of shifts in the
quasi-orders considered in this paper. We will use it later in
section~\ref{sec:lattice} but we now state the main result of this
section concerning levels at the bottom of the quasi-orders.

\begin{cor}
  \label{cor:infinishi}
  Let $\simu$ be a relation among $\sacsimu$, $\facsimu$ and
  $\facsacsimu$. For any ${n\geq3}$, there are infinitely many
  incomparable CA at level $n$ for $\simu$.
\end{cor}
\begin{pf}
  We have shown in theorem~\ref{thm:nilper} that translations CA are at
  level $1$. Lemma~\ref{lm:soushi} together with
  lemma~\ref{lm:canoshi} show that a product of two translations (with
  distinct vectors) is at level $2$. By theorem~\ref{thm:shishi} we
  conclude that any product of $n$ translations with distinct vectors
  is at level $n$ and two such CA are incomparable if they have
  different characteristic sequences provided ${n\geq3}$.\qed
\end{pf}

\section{Structural properties}
\label{sec:sp}

In this section we study in various ways the order structures induced
by the simulation relations defined above.

\subsection{Cartesian Products and Lack of (Semi-)Lattice Structure}
\label{sec:lattice}

The Cartesian Product of cellular automata is not a neutral operation
from the point of view of the three quasi-orders of the paper. For
instance, there are CA $\ACA$, $\ACB$ which are equivalent but such
that ${\ACA\times\ACA}$ is not equivalent to ${\ACB\times\ACA}$ (it is
sufficient to take two shifts with different translation vectors).

The next theorem shows however that some simulations by Cartesian
products of CA can be transposed to components of the product. 

\begin{thm}
  \label{thm:deprod}
  Let $\ACA$ be a CA with two states and let $\simu$ be a simulation
  relation among $\sacsimu$, $\facsimu$ and $\facsacsimu$. For any
  $\ACB$ and $\ACC$, if ${\ACB\times\ACC}$ strongly $\simu$-simulates
  $\ACA$ then either ${\ACA\simu\ACB}$ or ${\ACA\simu\ACC}$.
\end{thm}

Notice that the 'two states' and 'strong simulation' hypotheses are
both important and related. The theorem doesn't hold without such
hypotheses: take two shifts with different vectors for $\ACB$ and
$\ACC$ and choose $\ACA=\ACB\times\ACC$.

\begin{pf}[of the theorem]
  Let ${\alphA=\{a_1,a_2\}}$.  First, we show that
  ${\ACA\sac_\iota\ACB\times\ACC}$ implies either ${\ACA\sac\ACB}$ or
  ${\ACA\sac\ACC}$ which is sufficient to prove the theorem for
  $\sacsimu$ and $\facsacsimu$. Since ${\iota(a_1)\not=\iota(a_2)}$ we have
  either ${\pi_1(i(a_1))\not=\pi_1(i(a_2))}$ or
  ${\pi_2(i(a_1))\not=\pi_2(i(a_2))}$ where $\pi_1$ and $\pi_2$ are
  projections over first and second component respectively. We suppose
  the first case (the second is symmetric) and so ${\pi_1\circ i :
    \alphA\rightarrow\alphB}$ is injective. Moreover, since
  \[\unif{\pi_1\circ \iota}\circ\globA = \unif{\pi_1}\circ
  G_{\ACB\times\ACC}\circ\unif{\iota} = \globB\circ\unif{\pi_1\circ
    \iota},\] we conclude that ${\ACA\sac_{\pi_1\circ \iota}\ACB}$.

  Second, we show that ${\ACA\fac_s\ACB\times\ACC}$ implies either
  ${\ACA\facsimu\ACB}$ or ${\ACA\facsimu\ACC}$ which is sufficient to
  prove the theorem for $\facsimu$.  Let $I_\ACA$ be the set of states
  that can be reached after one step of $\ACA$ (formally,
  ${I_\ACA=\locA(\alphA,\ldots,\alphA)}$) and $I_\ACB$ and $I_\ACC$ be
  similar sets for $\ACB$ and $\ACC$.

  \begin{itemize}
  \item We first suppose that $\ACB$ and $\ACC$ are such that each
    uniform configuration is either a fixed-point or without any
    uniform antecedent. If ${a_1\not\in I_\ACA}$ then for any $b$ and
    $c$ such that ${s(b,c)=a_1}$ we have either ${b\not\in I_\ACB}$ or
    ${c\not\in I_\ACC}$. We suppose the first case (the second is
    analogous) and then we have ${\ACA\fac_\zeta\ACB}$ where
    ${\zeta:\alphB\rightarrow\alphA}$ is defined by ${\zeta(b) = a_1}$
    and ${\zeta(x)=a_2}$ for ${x\not=b}$.
    
    If ${a_2\not\in I_\ACA}$ we apply the same reasoning and so we
    are left with the case ${I_\ACA = \alphA}$. Since pairs of
    ${\alphB\times\alphC}$ are $2$-colored via $s$, there must be
    two pairs of different colors which agree on a
    component. Suppose it is the first component (the other case is
    symmetric), we have ${b_1,b_2\in\alphB}$ and ${c\in\alphC}$ such
    that ${s(b_1,c)=a_1}$ and ${s(b_2,c)=a_2}$. Consider the set
    ${X=\{(b_1,c),(b_2,c)\}}$. Since
    ${\unif{s}\circ(\globB\times\globC) = \globA\circ\unif{s}}$ and
    ${\unif{s}\bigl(X^\ZZ\bigr) = \alphA^\ZZ}$ and ${I_\ACA =
      \alphA}$ we necessarily have ${s(\alphB,d) = \alphA}$ where
    $d$ defined by ${\unif{d}=\globC(\unif{c})}$. $d$ is quiescent
    by hypothesis on $\ACC$. So we have ${\ACA\fac_\zeta\ACB}$ with
    ${\zeta:\alphB\rightarrow\alphA}$ defined by ${\zeta(x) =
      s(x,d)}$ ($\zeta$ is onto by choice of $d$).
  \item Now suppose that the hypothesis on $\ACB$ and $\ACC$ are not
    fulfilled. Then, if ${t = |\alphB|!\times|\alphC|!}$, both
    $\ACB^t$ and $\ACC^t$ are guarantied to fulfill the required
    hypothesis (because any uniform configuration is either in a cycle
    of uniform configurations, or without uniform antecedent
    arbitrarily far in the past). Since
    ${\ACA^t\fac\bigl(\ACB\times\ACC\bigr)^t=\ACB^t\times\ACC^t}$, it
    suffices to apply the previous reasoning on $\ACA^t$, $\ACB^t$ and
    $\ACC^t$ to conclude either ${\ACA^t\fac\ACB^t}$ or
    ${\ACA^t\fac\ACC^t}$. In either case the theorem follows.\qed
  \end{itemize}
\end{pf}

The Cartesian product operation is not a supremum in any of the
quasi-order.  In fact these quasi-orders don't admit any supremum or
infinimum operation as shown by the theorem below. Recall that an
upper semi-lattice is a partial order structure $\leq$ equipped with a
'sup' operation such that:
\[a\leq x\text{ and }b\leq x\ \Rightarrow\ \sup(a,b)\leq x\] The
definition for lower semi-lattice is dual (replace 'sup' by 'inf' and
any relation $x\leq y$ by $y\leq x$).

\begin{thm}
  \label{thm:noshishi}
  Let $\simu$ be a relation among $\sacsimu$, $\facsimu$ and
  $\facsacsimu$. Then the ordered structure ${\bigl(AC/\sime,
    \simu\bigr)}$ is neither an upper semi-lattice, nor a lower
  semi-lattice.
\end{thm}
\begin{pf}
  Let $\ACA_{2}$, $\ACA_{3}$, $\ACA_{2,3}$ and $\ACA_{2,4}$ be
  products of translations with characteristic sequences $(2)$, $(3)$,
  ${(2,3)}$ and ${(2,4)}$ respectively. Theorem~\ref{thm:shishi} and
  lemma~\ref{lm:canoshi} show that they induce the following structure
  in $\simu$:
  \begin{center}
    \begin{tikzpicture}
      \tikzstyle{every node}=[shape=circle];%
      \path (0,0) node[below] (v0) {$\ACA_2$};%
      \path (2,0) node[below] (v1) {$\ACA_3$};%
      \path (0,1) node[above] (v2) {$\ACA_{2,3}$};%
      \path (2,1) node[above] (v3) {$\ACA_{2,4}$};%
      \draw[->, thick, black] (v2) -- (v1);%
      \draw[->, thick, black] (v2) -- (v0);%
      \draw[->, thick, black] (v3) -- (v1);%
      \draw[->, thick, black] (v3) -- (v0);%
    \end{tikzpicture}
  \end{center}
  where an arrow from $\ACA$ to $\ACB$ means ${\ACB\simu\ACA}$ and if
  ${\ACB\simu\ACC\simu\ACA}$ then either ${\ACB\sime\ACC}$ or
  ${\ACC\sime\ACA}$. This shows that the pair $\ACA_{2}$, $\ACA_{3}$
  has no supremum and that the pair $\ACA_{2,3}$, $\ACA_{2,4}$ has no
  infimum.\qed
\end{pf}

\subsection{Ideals and Filters}
\label{sec:idfi}

Although the structures ${\bigl(AC,\simu\bigr)}$ studied in this paper
are not semi-lattices (see above), many classical properties of
cellular automata are nicely captured through \emph{ideals} and
\emph{filters}. Well-known in lattice theory and algebra, the notions
of ideal and filter can also by defined for an arbitrary
(quasi-)ordered structure \cite{lattice}. For the structure
${\bigl(AC,\simu\bigr)}$, an ideal $I$ is a set of CA such that:
\begin{itemize}
\item if ${\ACA\in I}$ and ${\ACB\simu\ACA}$ then ${\ACB\in I}$;
\item for any ${\ACA,\ACB\in I}$ there is some ${\ACC\in I}$ such that
  ${\ACA\simu\ACC}$ and ${\ACB\simu\ACC}$.
\end{itemize}
Moreover, $I$ is said \emph{principal} if there is some $\ACA_I$ such
that ${\ACA\in I\iff\ACA\simu\ACA_I}$.  The notion of filter and
principal filter are dual of ideal and principal ideal (replacing all
${X\simu Y}$ by ${Y\simu X}$).

Given a set $I$ of CA, the three following conditions are sufficient for
$I$ to be an ideal for the simulation $\sacsimu$ (resp. $\facsimu$, or
$\facsacsimu$):
\begin{enumerate}
\item ${\ACA\in I\iff\bulk{\ACA}{\alpha}\in I}$ for any transform $\alpha$,
\item if ${\ACB\in I}$ and ${\ACA\sac\ACB}$ (resp. ${\ACA\fac\ACB}$,
  or ${\ACA\facsac\ACB}$) then ${\ACA\in I}$,
\item if ${\ACA\in I}$ and ${\ACB\in I}$ then ${\ACA\times\ACB\in I}$.
\end{enumerate}

Most of the proofs below follow this scheme.  

\subsubsection{Dynamical properties}

The following theorem shows that several dynamical properties of
global rules of CA correspond to ideals in the quasi-orders.  A CA is
nilpotent over periodic configurations if there exists a spatially
periodic configuration $c_0$ such that all spatially periodic
configurations lead in finite time to $c_0$.

\begin{thm}
  \label{thm:commonideal}
  Let $\simu$ be a simulation relation among $\sacsimu$, $\facsimu$
  and $\facsacsimu$. The following sets of CA form ideals of
  ${\bigl(AC,\simu\bigr)}$:
  \begin{itemize}
  \item surjective CA,
  \item reversible CA,
  \item CA which are nilpotent over periodic configurations.
  \end{itemize}
\end{thm}
\begin{pf}
  First, from the point of view of global maps, a geometric transform consists
  in iterating or composing with bijective maps. So the properties of being
  surjective or reversible are left unchanged by geometrical
  transforms. Besides, geometrical transforms map periodic configurations to
  periodic configurations, cycles of configurations to cycles of configurations
  (possibly reduced to a single configuration), and attraction basins of such
  cycles to attraction basins of cycles. Hence, nilpotency over periodic
  configurations, which is equivalent to the existence of a temporal cycle having
  all periodic configurations in its attraction basin, is preserved by
  geometrical transforms. By similar reasoning on the phase space, it is
  straightforward to check that $\ACA$ is nilpotent over periodic configurations
  if $\ACB$ is and ${\ACA\sac\ACB}$ or ${\ACA\fac\ACB}$. And ${\ACA\times\ACB}$
  is nilpotent over periodic configurations if both $\ACA$ and $\ACB$ are. So
  nilpotency over periodic configurations induces an ideal for $\simu$.

  It is also clear that surjectivity and reversibility are preserved
  by cartesian product. Now suppose ${\ACA\fac_\pi\ACB}$. If $\ACB$ is
  surjective then so is $\ACA$ since
  ${\globA\circ\unif{\pi}=\unif{\pi}\circ\globB}$ and $\pi$ is by
  definition surjective. If $\ACB$ is reversible, consider any map
  $\phi$ such that ${\pi\circ\phi=Id}$ and let ${\ACA_{-1}}$ be the CA
  over state set $\alphA$ and defined by the global map ${G =
    \unif{\pi}\circ\globB^{-1}\circ\unif{\phi}}$ (it is a
  shift-commuting continuous map). Since
  ${\globA\circ\unif{\pi}=\unif{\pi}\circ\globB}$, one can check that
  ${\globA\circ G = Id}$ so $\ACA$ is reversible.
  
  Finally, suppose ${\ACA\sac_\iota\ACB}$.  If $\ACB$ is reversible
  then $\ACA$ is also reversible since ${\unif{\iota}\circ\globA =
    \globB\circ\unif{\iota}}$ and $\iota$ is by definition
  injective. If $\ACB$ is surjective, then so is $\ACA$ because $\ACB$
  being injective over finite configurations (Moore-Myhill
  theorem\footnote{In \cite{Hedlund:1969}, one can find the following
    theorem: a CA is surjective if and only if there is no pair of
    finite configurations (\textit{i.e.} uniform except on a finite
    region) having the same image. The original formulation of the
    Moore-Myhill theorem \cite{moore,myhill} supposes the existence of
    a quiescent state.})  $\ACA$ is also injective over finite
  configurations ($\iota$ maps finite configurations to finite
  configurations). \qed
\end{pf}

\begin{thm}
  Let $\ACA$ and $\ACB$ be two reversible CA and $\simu$ be a
  simulation relation among $\sacsimu$, $\facsimu$ and
  $\facsacsimu$. If ${\ACA\simu\ACB}$ then
  ${\ACA^{-1}\simu\ACB^{-1}}$.
\end{thm}
\begin{pf}
  First, it is straightforward to check that the inverse of
  geometrically transformed instances of $\ACA$ are transformed
  instances of the inverse of $\ACA$. Using what was shown above
  concerning reversibility, it is thus sufficient to prove the two
  following properties:
  \begin{itemize}
  \item ${\ACA\sac_\iota\ACB}$ implies ${\ACA^{-1}\sac_\iota\ACB^{-1}}$,
  \item ${\ACA\fac_g\ACB}$ implies ${\ACA^{-1}\fac_g\ACB^{-1}}$.
  \end{itemize}
  In the first case we have:
  \[\globB\circ\unif{\iota}=\unif{\iota}\circ\globA\Rightarrow \unif{\iota} =
  \globB^{-1}\circ\unif{\iota}\circ\globA\Rightarrow
  \unif{\iota}\circ\globA^{-1} = \globB^{-1}\circ\unif{\iota}\] each
  equality being true on ${\alphA^\ZZ}$.  In the second case we have:
  \[\globA\circ\unif{g}=\unif{g}\circ\globB\Rightarrow \globA\circ\unif{g}\circ\globB^{-1}=\unif{g}\Rightarrow \unif{g}\circ\globB^{-1} =
  \globA^{-1}\circ\unif{g}\] each equality being true on
  ${\alphB^\ZZ}$.  \qed
\end{pf}

One immediate consequence of the theorem is that if two reversible CA
are equivalent then their inverse CA are also equivalent. What is not
obvious however is whether the inverse CA are necessarily in the same
class as the initial CA.

\begin{openpb}
  Consider any simulation relation and $\sime$ the associated equivalence
  relation.  What are the reversible CA $F$ such that $F\sime F^{-1}$?
\end{openpb}

At the time of writing we have no example of a reversible $F$ with
$F\not\sime F^{-1}$. 

\begin{thm}
  \label{thm:revuniv}
  Let $\simu$ be $\sacsimu$ or $\facsacsimu$. Then the ideal of
  reversible CA is principal: there is a reversible CA $\ACA$ such
  that
  \[\ACB\text{ reversible}\iff\ACB\simu\ACA.\]
\end{thm}
\begin{pf}
  In \cite{DurandLose97}, a reversible CA $\ACB$ able to simulate any
  reversible CA is constructed. The notion of simulation used is
  included in $\sacsimu$ and therefore in $\facsacsimu$. The
  implication $\Rightarrow$ is thus proven and the converse
  implication is proven by theorem~\ref{thm:commonideal}.\qed
\end{pf}

For the ideal of surjective CA, the principality is still an open
problem in dimension 1.

\begin{openpb}
  \label{open:surjideal}
  Is the ideal of surjective CA principal, and for which simulation
  quasi-order?
\end{openpb}

Limit sets of CA have received a lot of attention in the literature
\cite{Culik89, hur87, goles93}. The limit set of $\ACA$ is the set
$\Omega_\ACA$ of configurations having predecessors arbitrarily far in
the past, formally:
\[\Omega_\ACA = \bigcap_t \globA^t\bigl(\alphA^\ZZ\bigr).\]

The next theorem shows that the class of CA with a sofic limit set is
nicely captured by $\facsimu$.

\begin{thm}
  \label{thm:soficideal}
  The set of CA with a sofic limit set is an ideal for $\facsimu$.
\end{thm}
\begin{pf}
  For CA of dimension 1, having a sofic limit set is equivalent to having a
  regular limit language \cite{weiss}. It is clear that this latter property is
  left unchanged by geometrical transforms (the limit language is not affected by
  iterations and shifts, the regularity of the language is not affected by
  packing). Hence, it is sufficient to show that if $\ACB$ has a regular limit
  language and ${\ACA\fac_g\ACB}$ then $\ACA$ also has a regular limit
  language. Since regular languages are closed under substitution (a classical
  result which can be found in \cite{HopcroftUllam}), it is sufficient to prove
  that ${\Omega_\ACA = \unif{g}\bigl(\Omega_\ACB\bigr)}$. This last assertion is
  a direct consequence of ${\ACA\fac_g\ACB}$, since the following equality holds
  by recurrence on $t$:
  \[\unif{g}\bigl(\globB^t(\alphB^\ZZ)\bigr) =
  \globA^t(\alphA^\ZZ).\]\qed
\end{pf}

\begin{openpb}
  Let $\simu$ be $\sacsimu$ or $\facsacsimu$. Is there a $\simu$-universal CA
  with a sofic limit set?
\end{openpb}

\subsubsection{Topological dynamics}

The properties considered above are purely dynamic: they can be
expressed as structural properties of the phase space with the
reachability relation only. We now consider properties from
topological dynamics: they are expressed with both the reachability
relation and the topology (Cantor distance) of the space of
configurations.  We will show that many of them correspond to ideals
of the simulation quasi-orders.

The properties we will consider are derived from the equicontinuity
classification of P.~K{\r u}rka \cite{Kurka97}. Let $\ACA$ be any CA
with state set $Q$ and global rule $G$ and denote by $\dist$ the
Cantor distance over $Q^\ZZ$.
\begin{itemize}
\item $x\in Q^\ZZ$ is an \emph{equicontinuity point} for $\ACA$ if 
  \[\forall\epsilon, \exists\delta, \forall y\in Q^\ZZ :
  \dist(x,y)\leq\delta\Rightarrow \forall t,
  d(G^t(x),G^t(y))\leq\epsilon.\]
\item $\ACA$ is \emph{sensitive to initial conditions} if
  \[\exists\epsilon, \forall\delta, \forall x\in Q^\ZZ\exists y\in Q^\ZZ
  \exists t : \dist(x,y)\leq\delta\text{ and
  }\dist\bigl(G^t(x),G^t(y)\bigr)\geq\epsilon.\]
\item $\ACA$ is \emph{(positively) expansive} if
  \[\exists\epsilon,\forall x,y\in Q^\ZZ : x=y\iff\forall t,
  \dist\bigl(G^t(x),G^t(y)\bigr)\leq\epsilon.\]
\end{itemize}

The classification of P.~K{\r u}rka is the following:
\begin{description}
\item[$K_1$] is the set of CA for which all configurations are equicontinuity points,
\item[$K_2$] is the set of CA having equicontinuity points,
\item[$K_3$] is the set of CA sensitive to initial conditions,
\item[$K_4$] is the set of expansive CA.
\end{description}

The weakness of this classification is its lack of shift-invariance:
the identity and the elementary translation belong to different
classes ($K_1$ and $K_3$ respectively). Several attempts have been
made to overcome this problem by changing the topology \cite{besico}.
More recently, a new approach has been proposed \cite{sablikTCS}: the
Cantor topology is conserved (with all its good properties) but the
topological properties are enriched with a new parameter (a velocity)
which is used as the reference direction of information propagation in
space-time. The original definitions of P.~K{\r u}rka are thus
obtained by choosing velocity $0$, but now identity and elementary
translations are assigned to the same class (with different
velocities). This directional dynamic approach is more suitable for
our study since, by definition, the equivalence classes of any of our
quasi-orders are shift-invariant. We will define $4$ classes based on
the existence of some direction for which some dynamical behavior is
observed.

We say that $\ACA$ is a \emph{rescaling} of $\ACB$ if there are
transforms $\alpha$ and $\beta$ such that
${\bulk{\ACA}{\alpha}\isom\bulk{\ACB}{\beta}}$.  We then consider the
following $4$ classes:

\begin{itemize}
\item the set $\equi$ of CA which are a rescaling of some equicontinuous CA,
\item the set $\equipt$ of CA which are a rescaling of some CA having
  equicontinuity points,
\item the set $\sensi$ of CA which are not in $\equipt$, \textit{i.e.}
  CA which are sensitive in every directions\footnote{For
    one-dimensional CA, the set of sensitive CA is the complement of
    the set of CA having equicontinuity points (see \cite{Kurka97}). In
    \cite{sablikTCS}, this complementarity is shown for any
    direction.},
\item the set $\expansi$ of CA which are a rescaling of some
  (positively) expansive CA.
\end{itemize}

\begin{figure}[htbp]
  \centering
  \begin{tikzpicture}
    \draw (0,0) ellipse (2.5 cm and 1.5 cm);%
    \path (-1.1,-.8) node () {$\equipt$};%
    \path (1.1,.8) node () {$\sensi$};%
    \filldraw[fill=black!10, draw=black] (-1,.5) node (u) {$\equi$} ellipse (.7cm and .5cm);
    \filldraw[fill=black!10, draw=black] (1,-.5) node (v) {$\expansi$} ellipse (.7cm and .5cm);
    \draw (0,1.5).. controls (.5,0) and (-.5,0)..(0,-1.5);
    \path (-3.5,1)  node[shape=circle] (id) {$Id$};
    \path (-3.5,-1)  node[shape=circle] (max) {$\delta_{\max}$};
    \path (4,1)  node[shape=circle] (sig) {$\sigma_1\times\sigma_{-1}$};
    \path (3.5,-1)  node[shape=circle] (ztwo) {$\zee{2}$};
    \draw [->, thick] (id) -- (-1.4,.5);
    \draw [->, thick] (max) -- (-2,-.3);
    \draw [->, thick] (sig) -- (1.8,.3);
    \draw [->, thick] (ztwo) -- (1.4,-.5);
  \end{tikzpicture}
  \caption{Four kinds of topological dynamics.}
  \label{fig:shiftkurka}
\end{figure}
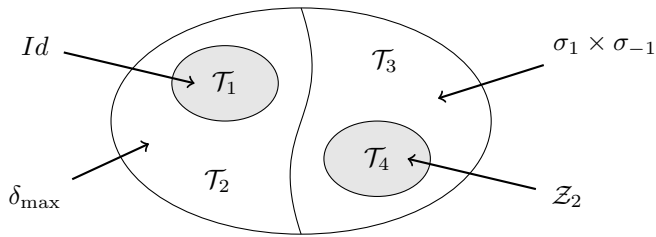

Figure~\ref{fig:shiftkurka} is justified by the following theorem.

\begin{thm} 
  We have the following inclusions:
  \begin{enumerate}
  \item ${\equi\subseteq\equipt}$,
  \item ${\expansi\subseteq\sensi}$.
  \end{enumerate}
  Moreover, each of the sets $\equi$, $\equipt\setminus\equi$,
  $\sensi\setminus\expansi$ and $\expansi$ is non-empty.
\end{thm}
\begin{pf}
  The first inclusion follows from definitions. The second follows
  from proposition 3.2 of \cite{sablikTCS} which asserts that the set
  of directions with equicontinuity points and the set of expansive
  directions cannot be simultaneously empty.

  Non-emptyness of $\equi$ and $\expansi$ follows from the existence of
  equicontinuous (\textit{e.g.} the identity) and (positively)
  expansive CA (\textit{e.g.} $\zee{2}$). Moreover, any CA having an
  equicontinuity point which is not equicontinuous (\textit{e.g.} the
  CA of local rule ${\delta_{\max}(a,b,c)=\max(a,b,c)}$) cannot be in
  $\equi$ (equicontinuity is preserved by rescaling), so it is in
  ${\equipt\setminus\equi}$. Finally,
  ${\sigma_1\times\sigma_{-1}\in\sensi\setminus\expansi}$. Indeed, for
  $\sigma\times\sigma^{-1}$, any direction is either a direction of
  right-expansivity or a direction of left-expansivity, neither
  both. So $\sigma\times\sigma^{-1}\not\in\expansi$. Finally,
  $\sigma\times\sigma^{-1}\not\in\equipt$ since, by proposition 3.2 of
  \cite{sablikTCS}, no direction of (left or right) expansivity can be
  a direction with equicontinuity points.\qed
\end{pf}

\begin{thm}
  \par\noindent
  \label{thm:tdideals}
  \begin{enumerate}
  \item $\equi$ is an ideal for any simulation $\simu$ among
    $\facsimu$, $\sacsimu$ and $\facsacsimu$;
  \item $\equipt$ is an ideal for $\facsimu$;
  \item ${\expansi}$ is an ideal for $\sacsimu$.
  \end{enumerate}
\end{thm}
\begin{pf}
  First, consider any ${\ACA,\ACB\in\equi}$. Then there are CA $\ACA'$
  and $\ACB'$ which are both equicontinuous and $\simu$-equivalent to
  $\ACA$ and $\ACB$ respectively. Then, if ${\ACC=\ACA'\times\ACB'}$
  we have ${\ACC\in\equi}$ and by theorem~\ref{thm:bfacts} we have
  both ${\ACA\simu\ACC}$ and ${\ACB\simu\ACC}$. The same reasoning can
  be applied to $\expansi$ and $\equipt$. Thus we have shown the
  second condition of the definition of ideals for the three properties
  considered here.

  To conclude the theorem, and since the three properties considered are
  by definition invariant by rescaling, it is sufficient to prove:
  \begin{itemize}
  \item if ${\ACA\sac\ACB}$ or ${\ACA\fac\ACB}$ then ${\ACB\text{ equicontinuous}\Rightarrow\ACA\text{ equicontinuous}}$;
  \item if ${\ACA\fac\ACB}$ then ${\ACB\text{ has equicontinuous points}\Rightarrow\ACA\text{ has equicontinuous points}}$;
  \item if ${\ACA\sac\ACB}$ then ${\ACB\text{
        expansive}\Rightarrow\ACA\text{ expansive}}$.
  \end{itemize}
  
  The first assertion follows from the characterisation of
  equicontinuous CA as ultimately periodic CA \cite{Kurka97}. 
  
  For the second assertion, if ${\ACA\fac_\pi\ACB}$ then for all
  ${x,y\in\alphB^\ZZ}$ we have the inequality
  ${\dist(\pi(x),\pi(y))\leq\dist(x,y)}$.  Moreover, for any
  ${y_1\in\alphA^\ZZ}$ there is some $y_2\in\alphB^\ZZ$ such that
  ${\pi(y_2)=y_1}$ and ${\dist(\pi(x),y_1) = \dist(x,y_2)}$ (choose
  $y_2$ so that it equals $x$ on the cells around position $0$).
  Hence, if $x$ is an equicontinuous point for $\ACB$ then $\pi(x)$ is
  an equicontinuous point for $\ACA$.

  Finally, for the third assertion, it is sufficient to notice that
  the property of expansivity is defined by a formula using only
  universal quantifications on configurations so it remains true on a
  subset of configurations. \qed
\end{pf}

$\equipt$ is not an ideal for $\sacsimu$ and neither for
$\facsacsimu$ as shown by the following example.
\begin{ex}
  Consider $\ACB\in\sensi$ of radius $1$ and let $\ACA$ be the CA with
  radius $1$, states set ${\alphA=\alphB\cup\{M\}}$ (with ${M\not\in
    \alphB}$) with local rule $\locA$ defined by
  \[\locA(x,y,z) =
  \begin{cases}
    \locB(x,y,z)&\text{if }\{x,y,z\}\subseteq \alphB,\\
    y&\text{else.}
  \end{cases}\] ${\ACB\sac\ACA}$ so ${\ACB\sacsimu\ACA}$. However
  ${\ACA\in\equipt}$ since the configuration ${{}^\omega M^\omega}$ is
  an equicontinuous point.\qed
\end{ex}

Notice also that $\sensi$ cannot be an ideal because
${\sigma_1\times\sigma_{-1}\in\sensi}$ simulates ${\sigma\in\equi}$.

\begin{openpb}
  \label{open:expansif}
  Are there ${\ACA\not\in\expansi}$ and ${\ACB\in\expansi}$ such that
  ${\ACA\facsimu\ACB}$ (\textit{i.e.} the simulator CA is expansive up
  to rescaling but the simulated CA is not expansive, even up to
  rescaling)?
\end{openpb}

\subsection{(Un)decidability}
\label{sec:undec}

The fact that many properties related to the simulation quasi-orders are
undecidable is no surprise. For instance the nilpotency property, which is an
undecidable problem \cite{kari92}, corresponds to an equivalence class in the
three quasi-orders (theorem~\ref{thm:nilper}). However, there are non-trivial
properties of these quasi-orders which are decidable (see below) and the edge
between decidable and undecidable properties is hard to catch.

In this section, we consider two kinds of problems in simulation
quasi-orders: lower bounds (being above some fixed CA or set of CA)
and upper bounds (being simulated by some fixed CA or some CA from a
fixed set).

\begin{thm}[\cite{mazrap}]
  The set of CA of radius $1$ with a spreading state and nilpotent
  over periodic configurations is not co-recursively enumerable.
\end{thm}

\begin{thm}
  \label{thm:nilperdeci}
  Let $\ACA$ be any CA which is not nilpotent over periodic
  configurations. Let $\simu$ be either $\sacsimu$ or
  $\facsacsimu$. Then the set of CA $\ACB$ such that ${\ACA\simu\ACB}$
  is not co-recursively enumerable.
\end{thm}
\begin{pf}
  We describe a computable transformation which, given a CA $\ACC$ of radius $1$
  with a spreading state, produces a CA $\ACB$ with the following properties:
  \begin{itemize}
  \item if $\ACC$ is not nilpotent over periodic configurations then
    ${\ACA\simu\ACB}$;
  \item if $\ACC$ is nilpotent over periodic configurations then so is
    $\ACB$.
  \end{itemize}
  The theorem follows by theorem~\ref{thm:commonideal} since we have
  reduced the problem '${\ACA\simu\ACB}$?' to the problem of
  nilpotency over periodic configurations (reduced to CA of radius $1$
  with a spreading state).

  We now describe the construction of $\ACB$ from $\ACC$.  Suppose
  $\ACC$ has a spreading state $q$. $\ACB$ is the CA of radius $1$ and
  states set ${\alphB = (\alphC\setminus\{q\})\times\alphA\cup\{q\}}$
  with local rule $\locB$ defined by:
  \[\locB(a,b,c) =
  \begin{cases}
    \bigr(\locC(a_1,b_1,c_1),\locA(a_2,b_2,c_2)\bigr)
    &\text{ if }
    \begin{cases}
      a,b,c\in\alphB\setminus\{q\} \text{ and}\\
      \locC(a_1,b_1,c_1)\not=q,
    \end{cases}
    \\
    q&\text{ in any other case,}
  \end{cases}
  \]
  where $a_i$, $b_i$ and $c_i$ represent component $i$ of $a$, $b$ and
  $c$. Any periodic configuration $c$ of $\ACB$ either leads to the
  uniform configuration $\unif{q}$, or contains a periodic
  configuration of $\ACC$ in its first component. Hence, if $\ACC$ is
  nilpotent over periodic configurations, then so is $\ACB$ (because
  $q$ is precisely the spreading state of $\ACC$).  If $\ACC$ is not
  nilpotent over periodic configurations, then there is a word
  ${u\in(\alphC\setminus\{q\})^m}$ and an integer ${t\geq 1}$ such
  that the periodic configuration $c$ of period $u$ verifies
  ${\globC^t(c) = c}$. Therefore, by definition of $\ACB$, we have
  ${\bulk{\ACA}{m,1,t,0}\sac_i\bulk{\ACB}{m,1,t,0}}$ where
  ${i:\alphA^m\rightarrow\alphB}$ is defined by:
  \[i(a_1,a_2,\ldots, a_m) =
  \bigl((a_1,u_1),\ldots,(a_m,u_m)\bigr).\]\qed
\end{pf}

This result shows that it is generally undecidable to know whether a
CA is lower-bounded by a given (fixed) one. However, there are
noticeable exceptions in one dimension for $\facsimu$ and $\facsacsimu$.

\begin{thm}
  \label{thm:abovenilpo}
  Let $\simu$ be either $\facsimu$ or $\facsacsimu$ and let $\ACA$ be
  a nilpotent CA. Then the problem of determining if a given $\ACB$ is
  above $\ACA$ for $\simu$ is decidable.
\end{thm}
\begin{pf}
  We are going to show that ${\ACA\simu\ACB}$ if and only if $\ACB$ is
  not surjective and the theorem follows by decidability of
  surjectivity in one dimension \cite{amoroso72}.  First, by
  theorem~\ref{thm:commonideal}, if $\ACB$ is surjective then
  ${\ACA\cancel\simu\ACB}$. Suppose now that $\ACB$ is not surjective,
  \textit{i.e.} that $\ACB$ possesses some Eden word ${u\in\alphB^m}$
  for some length $m$. Then, denoting by $\ACC$ the CA over states set
  ${\alphC=\{0,1\}}$ which is constant equal to $0$, we have
  ${\ACC\fac_\pi\bulk{\ACB}{m,1,1,0}}$ if
  ${\pi:\alphB^m\rightarrow\alphA}$ verifies ${\pi(w) = 0}$ if and
  only if ${w=u}$. We deduce by theorem~\ref{thm:nilper} that
  ${\ACA\simu\ACB}$.\qed
\end{pf}

\begin{openpb}
  \label{open:abovenilpo}
  Is there a non-surjective CA $\ACA$ which cannot injectively
  simulate any nilpotent CA? Is the problem of being above the class
  of nilpotent CA for injective simulation a decidable problem?
\end{openpb}

Concerning upper-bound problems, the edge between decidability and
undecidability is also non-trivial.  For instance,
theorem~\ref{thm:revuniv} shows the existence of a CA $\ACA$ such that
the upper-bound decision problem '${\ACB\simu\ACA}$?' is decidable in
dimension 1.

\section{Tops of the Orders}
\label{sec:top}

In this section we study maximal elements of the
quasi-orders. These CA are able to simulate any other CA

\begin{defn}
  Let $\simu$ be any relation among $\sacsimu$, $\facsimu$ and
  $\facsacsimu$. A CA $\ACA$ is said \emph{$\simu$-universal} if for
  any $\ACB$ we have ${\ACB\simu\ACA}$. It is \emph{strongly
    $\simu$-universal} if it strongly $\simu$-simulates any other CA.
\end{defn}

The notion of strong $\sacsimu$-universality above is exactly the same
notion as \emph{intrinsic universality} defined in section 5 of
\cite{bulking1} and has already been considered several times in the
literature (see \cite{JACsurvey} for a survey). In fact, strong and
general universality are the same notion for $\sacsimu$ and
$\facsacsimu$.

\begin{thm}
  \label{thm:stronguniv}
  There exist strongly $\sacsimu$-universal CA and all
  $\sacsimu$-universal CA are strongly $\sacsimu$-universal. The same
  is true for $\facsacsimu$.
\end{thm}
\begin{pf}
  For the existence of strongly $\sacsimu$-universal CA, see
  \cite{JACsurvey}. The theorem follows by application of
  theorem~12 of \cite{bulking1}. \qed
\end{pf}

Of course, any $\sacsimu$-universal is also
$\facsacsimu$-universal. The converse is an open problem.

\begin{openpb}
  Do the notions of $\sacsimu$-universality and
  $\facsacsimu$-universality coincide?
\end{openpb}

Concerning $\facsimu$, the situation is different: no CA is strongly
$\facsimu$-universal\footnote{The proof of this fact was suggested by
  G.~Richard.}.

\begin{thm}
  There is no strongly $\facsimu$-universal CA.
\end{thm}
\begin{pf}
  Suppose for the sake of contradiction that there is some strongly
  $\facsimu$-universal $\ACA$.  Consider a uniform configuration $c$
  of $\ACA$. There is $n$ such that the orbit of $c$ under $\ACA$
  contains $n$ different configurations (the orbit is ultimately
  periodic).  Now consider $\ACB$ with ${n+1}$ states such that its
  uniform configurations are all in the same cycle of length ${n+1}$.
  By hypothesis, for any $\ACB$ there is some geometric transform
  $\alpha$ such that ${\ACB\fac_s\bulk{\ACA}{\alpha}}$.  Let $d$ be
  the corresponding configuration of $c$ for
  $\bulk{\ACA}{\alpha}$. The orbit of $d$ contains at most $n$
  different configurations and it is therefore the same for the orbit
  of $s(d)$ under $\ACB$. But $s(d)$ is necessarily uniform and we get
  a contradiction with the choice of $\ACB$. \qed
\end{pf}

The theorem 12 of \cite{bulking1} don't apply for $\facsimu$. However,
we are not able either to construct a $\facsimu$-universal CA, nor to
prove that there is none.

\begin{openpb}
  \label{open:surjuniv}
  Is there a $\facsimu$-universal CA?
\end{openpb}

For the rest of this section, we consider only $\sacsimu$ and
$\facsacsimu$.

\subsection{On the Way to the Top}
\label{sec:reachtop}

Universal CA are not hard to construct and the property of being
universal is recursively enumerable since simulation relations
considered here are recursively enumerable. However universality is
not co-recursively enumerable as shown by the following theorem. The
case of $\sacsimu$-universality was proven in \cite{Ollinger03}. Using
theorem~\ref{thm:nilperdeci}, the proof below is direct and includes
the case of $\facsacsimu$.

\begin{thm}
  \label{thm:indeciuni}
  The set of $\sacsimu$-universal CA is not co-recursively enumerable
  and neither is the set of $\facsacsimu$-universal CA.
\end{thm}
\begin{pf}
  There exists a CA which is $\sacsimu$-universal but not nilpotent
  over periodic configurations. To see this consider any universal CA
  and add a new state which is spreading: the resulting CA, say
  $\ACA$, contains at least two disjoint periodic orbits of periodic
  configurations and is thus not nilpotent over periodic
  configurations. The theorem follows by application of
  theorem~\ref{thm:nilperdeci} to $\ACA$ since $\ACA$ is by
  construction both $\sacsimu$-universal and
  $\facsacsimu$-universal.\qed
\end{pf}

This result has some consequences on the structure of simulation
quasi-orders 'near' the top. The following theorem shows that a
non-universal CA is always 'infinitely far' from the class of
universal ones.

\begin{thm}
  \label{thm:notuniv}
  Let $\simu$ be $\sacsimu$ or $\facsacsimu$. And let $\univ$ be the
  set of $\simu$-universal CA. Then we have:
  \begin{enumerate}
  \item ${\ACA\times\ACB\in\univ\iff\ACA\in\univ\text{ or }\ACB\in\univ}$,
  \item if $\ACA\not\in\univ$ then there is $\ACB\not\in\univ$ with
${\ACA\simu\ACB}$ but ${\ACB\cancel\simu\ACA}$.
  \end{enumerate}
\end{thm}
\begin{pf}
  \par\noindent
  \begin{enumerate}
  \item By theorem~\ref{thm:bfacts} we have
    ${\ACA\simu\ACA\times\ACB}$ and ${\ACB\simu\ACA\times\ACB}$ which
    proves one direction. Moreover, there exists ${\ACC\in\univ}$ with
    $2$ states only \cite{banks,nicolasFCT}. If we suppose
    ${\ACA\times\ACB\in\univ}$ then, by theorem~\ref{thm:stronguniv},
    it strongly simulates $\ACC$. Hence, by theorem~\ref{thm:deprod},
    we have either ${\ACC\simu\ACA}$ or ${\ACC\simu\ACB}$ and thus
    either ${\ACA\in\univ}$ or ${\ACB\in\univ}$.
  \item Let ${\ACA\not\in\univ}$. If $\ACA$ was such that
    ${\ACC\simu\ACA}$ for all ${\ACC\not\in\univ}$ then the complement
    of $\univ$ would be the set ${\{\ACC : \ACC\simu\ACA\}}$ and
    $\univ$ would be co-recursively enumerable contradicting
    theorem~\ref{thm:indeciuni}. So there is ${\ACC\not\in\univ}$ with
    ${\ACC\cancel\simu\ACA}$. To conclude the proof it is sufficient
    to choose ${\ACB=\ACA\times\ACC}$ (theorem~\ref{thm:bfacts}).\qed
  \end{enumerate}
\end{pf}

\subsection{Necessary But Not Sufficient Conditions}
\label{sec:notuniv}

The purpose of this section is twofold. It compares the notions of
universality defined above to other definitions of the literature and,
by doing this, presents tools and techniques to prove non-universality
of some CA (other proofs of non-universality for other purposes are
developped in section~\ref{sec:io}). 

One of the techniques we use to ensure that some CA is not universal
yet achieving some behavior $B$, is to add a spreading state and let
the CA generates this state if it detects somewhere that the current
configuration doesn't correspond to a 'legal' configuration,
\textit{i.e.} a configuration occuring normally when producing the
behavior $B$.  Proofs of non-universality with this technique rely on
the lemma below. Before stating and proving the lemma, we need to give
some precisions on spreading states and sets of configurations
'supporting' a simulation.

First, the notion of spreading state is sensitive to the choice of the
syntactical representation of the CA because it depends on the choice
of the neihborhood. In the sequel we say a CA $\ACA$ has a spreading
state $\kappa$ if any cell changes to state $\kappa$ when $\kappa$
appears in its minimal neighborhood (\textit{i.e.} the minimal set of
cells upon which the local rule effectively depends).

Second, given a relation of the form
${\ACA\sac_i\bulk{\ACB}{m,1,t,z}}$, there is an isomorphism between
${(\ACA,\alphA^\ZZ)}$ and ${\bigl(\bulk{\ACB}{m,1,t,z},
  (i(\alphA))^\ZZ\bigr)}$ as dynamical systems. At the level of
$\ACB$, the configurations involved in this relation is the set $X$ of
configurations made of infinite concatenation of elements of
${i(\alphA)\subseteq\alphB^m}$ (viewed as words of length $m$ over
alphabet $\alphB$). This kind of sets are called \emph{block-subshifts}
and discussed in more details in section 3.2 of \cite{bulking1}. In
the sequel, such a set $X$ is called the \emph{support} of the
simulation.

\begin{lm}
  \label{lm:nospread}
  Let $\ACA$ be a CA without spreading state and $\ACB$ be a CA with a
  spreading state $\kappa$. If $\ACB$ strongly $\facsacsimu$-simulates
  $\ACA$, then the support $X$ of the simulation cannot contain
  $\kappa$.
\end{lm}
\begin{pf}
  By hypothesis, there are parameters $m$, $t$, $\tau$, $z$ and a CA $\ACC$ such that
  \[\ACA\fac_\pi \ACC\sac_i\bulk{\ACB}{m,\tau,t,z}.\]
  By choice of $\ACB$, $\bulk{\ACB}{m,\tau,t}$ admits $\kappa^m$ as
  spreading state. Moreover, by definition of $\facsac$, the minimal
  neighborhood of $\ACA$ is included in the minimal neighborhood of
  $\bulk{\ACB}{m,\tau,t}$. Thus, if $\kappa$ appears in some
  configuration of $X$ then the state ${\pi(i^{-1}(\kappa^m))}$ is a
  spreading state for $\ACA$ because $\kappa^n$ also appears in $X$
  for arbitrarily large $n$.\qed
\end{pf}

We first study how embeddings of Turing machines into CA can relate
the notions of universality for Turing machines to the notions of
universality derived from quasi-orders as defined above.

An embedding of a Turing machine $\MTM$ into a CA $\ACA$ is an
embedding of the instantaneous descriptions of $\MTM$ into
configurations of $\ACA$ such that instantaneous descriptions of
successive steps of $\MTM$ corresponds to successive steps of $\ACA$
via the embedding. We don't give any formal definition of embedding
since we will never prove negative results (\textit{i.e.} assertions of
the form `there is no embedding of $\MTM$ such that...'). However, the
embeddings we use in the sequel are classical and already appeared in
the literature (see \cite{Sutner03}).

\begin{thm}
  \label{thm:embed}
  For any Turing machine $\MTM$, there exists a CA $\ACA$ which embeds
  $\MTM$ but is not $\facsacsimu$-universal.
\end{thm}
\begin{pf}
  Let $\MTM=(\statM,\rubM,\transM)$ where $\statM$ is the set of
  states of $\MTM$, $\rubM$ is the tape alphabet, and \[\transM :
  \statM\times\rubM\rightarrow\statM\times\rubM\times\{-1,0,1\}\] is
  the transition function of $\MTM$.  We construct a CA $\ACA$ over
  state set \[\alphA = \rubM\times\{\leftarrow,\rightarrow\}\cup
  \rubM\times\statM\cup\{\kappa\}\] where $\rightarrow$ and
  $\leftarrow$ are states not already in $\statM$. Each cell of $\ACA$
  corresponds to a tape position of $\MTM$: it contains a letter from
  the tape alphabet and either a head with its current state or no
  head but an indication $\leftarrow$ or $\rightarrow$ telling in
  which direction to find the head. On configurations containing a
  single head, $\ACA$ mimics transitions of $\MTM$ step by step as
  expected. Thus, $\ACA$ embeds $\MTM$. In addition, $\ACA$ checks
  that $\leftarrow$ a never occur to the left of a state from $\statM$
  or a $\rightarrow$ (and symmetrically for $\rightarrow$). If the
  check fails, then the state $\kappa$ is generated and spreads.

  This construction ensures that, for any initial configuration $c$,
  if the orbit of $c$ never contains an occurrence of $\kappa$ then it
  contains at most one head. Hence, these orbits are such that at any
  time step state changes occur on the neighborhood of at most one
  position (a head move involves a state change in two adjacent
  cells).

  Now suppose that $\ACA$ is $\facsacsimu$-universal and consider the
  CA ${\ACB=\sigma_1\times\sigma_{-1}}$.  $\ACA$ strongly simulates
  $\ACB$ (theorem~\ref{thm:stronguniv}).  Since $\ACB$ has no
  spreading state, then the set $X$ of configurations of $\ACA$ on
  which the simulation occurs never contains $\kappa$. We deduce that
  all orbits of configurations from $X$ have the property described
  above. This implied that $\ACB$ is such that on all its orbits, at
  most two cells change their states between two steps: this in
  contradiction with the choice of $\ACB$.\qed
\end{pf}

Turing-universality of cellular automata is a fairly vague notion in
the literature. We don't give a formal definition here since we won't
prove any negative result concerning Turing-universality. We just
consider that a CA able to embed a universal Turing
machine\footnote{We don't give any formal notion of universality for
  Turing machine either. In fact, we only need to suppose the
  existence of at least one universal Turing machine.} is
Turing-universal.

We can choose $\MTM$ to be universal in the previous theorem
(theorem~\ref{thm:embed}). In this case, since the embedding used in
the proof ensures that $\MTM$ is simulated in real time by $\ACA$, we
deduce that the following problem is P-complete:
\begin{description}
\item[Input:] a state ${q\in\alphA}$, an integer ${t\geq 1}$, and a
  word ${u\in\alphA^{2rt+1}}$ where $r$ is the radius of $\ACA$;
\item[Query:] do we have ${\ACA^t(u)=q}$?
\end{description}
This problem of finite triangle computation has been considered
several times in the literature and it has been proven that it was
P-complete for particular CA \cite{GriffeathMoore, NearyW06}.  This
notion of complexity inherited from sequential computation theory
fails to capture the notion of universality associated to simulation
quasi-orders.

\begin{cor}
  There exists a CA which is Turing-universal and P-complete but not
  $\facsacsimu$-universal.
\end{cor}

\section{Induced Orders}
\label{sec:io}

This section aims at studying particular CA or sets of CA for the
ordered structure they induce in the simulation quasi-orders.  While
studying various properties of the quasi-orders in the previous
sections, we have already established the existence of several induced
infinite structures.

For instance, theorem~\ref{thm:notuniv} allows to construct an
infinite strictly increasing chain of non-universal CA starting from
any non-universal CA for the quasi-orders associated to $\sacsimu$ and
$\facsacsimu$.  Besides, theorem~\ref{thm:shishi} implies the
existence of inifinite chains in the three quasi-orders studied in this
paper.

Section~\ref{sec:limprod} below gives a way to construct chains of
length ${\omega+\omega}$ and an hint about the existence of chains of
length ${\omega\times\omega}$. However, we leave open the question of
the longest chain induced in any of the quasi-orders. We don't even
know if one of them admits a dense chain.

\begin{openpb}
  Does one of the quasi-orders admit a dense induced order?
\end{openpb}

\subsection{Limit Cartesian Product}
\label{sec:limprod}

We have seen in theorem~\ref{thm:notuniv} that if $\ACA$ is not
universal, then $\ACA\times\ACA$ cannot be universal. Therefore, no
finite Cartesian product of $\ACA$ with itself can be
universal. Therefore, we have a chain of non-universal CA:
\[\ACA\simu\ACA\times\ACA\simu\ACA\times\ACA\times\ACA\simu\cdots\]
For some $\ACA$, the chain collapses in a single equivalence class,
\textit{e.g.} if $\ACA$ is a translation (see
lemma~\ref{lm:canoshi}). However, the following theorem shows that for
some $\ACA$, the chain is strictly increasing. Moreover, $\ACA$ can be
chosen so that it embeds any Turing machine. 

\begin{thm}
  \label{thm:turingheads}
  For any Turing machine $\MTM$, there is a CA $\ACA$ which embeds
  $\MTM$ and such that for any ${1\leq n<m}$, one has:
  \[\underbrace{\ACA\times\cdots\times\ACA}_{m}\ \ \cancel\facsacsimu\ \ \underbrace{\ACA\times\cdots\times\ACA}_{n}.\]
\end{thm}
\begin{pf}
  Le $\ACA$ be the CA constructed in the proof of
  theorem~\ref{thm:embed}. We can suppose that $\MTM$ is such that it
  can produce infinite sequences of left move of its head when started
  from a special state (not the initial state) over a blank tape, and
  more precisely that the sequence of moves leaves the tape blank.  If
  $\MTM$ does not have this property, just add some states to achieve
  this behavior. We can suppose the same for right moves.
  
  Denote by $\ACB_m$ the product of $m$ copies of $\ACA$ and by
  $\ACB_n$ the product of $n$ copies. Suppose for the sake of
  contradiction that ${\ACB_m\facsacsimu\ACB_nq}$. We can construct
  for any set of positions ${z_1,\ldots,z_m}$ a configuration $c$ of
  $\ACB_m$ such that for all $i$ the $i$th component contains a
  correct instantaneous description of $\MTM$ where the head is at
  position $z_i$ in a state suitable to generate an infinite sequence
  of left or right moves (as supposed above).  Now let $c'$ be a
  configuration of $\ACB_n$ corresponding to $c$ via
  simulation. First, if some component $i$ of $c'$ contains a
  spreading state, it will spread and, after some time $t$, will be
  present at some position where the configuration
  $\glob{\ACB_m}^t(c)$ contains no head, but only a blank tape symbol
  on each component. This means that blocks of blank tape symbols in
  $\ACB_m$ can be simulated by blocks of $\ACB_n$ where the $i$th
  component is a block of spreading states. Considering again the
  orbit of $c$, we deduce that it can be simulated by a configuration
  $c''$ where the $i$th component is everywhere a spreading state
  except at a finite number of positions. Thus after some time, the
  $i$th component will become uniform and constant. It is then
  straightforward to show that it is useless for the simulation and
  that in fact $\ACB_m$ on $c$ can be simulated by only ${n-1}$ copies
  of $\ACA$.

  Applying the reasoning inductively, we can therefore suppose that no
  spreading state appears on any component in the orbit of the
  configuration $c'$ defined above. Since, the orbit of $c$ is such
  that there are $m$ distant positions where some states change at each
  step, it must be the case in the orbit of $c'$. Since, ${n<m}$,
  there must be some component with two heads and therefore a
  spreading state must appear after the first step: this is in
  contradiction with what we have just supposed.\qed
\end{pf}

For the CA $\ACA$ of the previous theorem, we can ask if the infinite
chain of Cartesian products is upper-bounded by some non-universal CA, or
if any CA able to simulate each product of the chain is necessarily
universal. One can imagine that for a sufficiently simple $\ACA$,
there is some room above the chain of products of $\ACA$ and below the
class of universal CA.

The rest of this section is devoted to the proof of a stronger result:
for any $\ACA$, there is a CA $\ACB$ which is able to simulate any
finite product of $\ACA$ and such that $\ACB$ is universal if and only
if $\ACA$ is universal. Moreover, $\ACB$ can be obtained from $\ACA$
constructively. Because it extends property of Cartesian product given
by theorem~\ref{thm:notuniv}, this construction will be called
\emph{limit product} in the sequel. If $\ACA$ is a CA, its limit
product is denoted by $\limprod{\ACA}$.

\textit{Note: }In the rest of this section we only consider the
simulation $\facsacsimu$.

Without loss of generality, we can suppose that $\ACA$ has radius $1$
(theorem~\ref{thm:bfacts}). To be able to simulate the product $\ACB$
of $n$ copies of $\ACA$, $\limprod{\ACA}$ is made of three layers (its
state set is a Cartesian product union a single state, which is a
spreading state as explain hereafter):
  \begin{enumerate}
  \item the \emph{state} layer,
  \item the \emph{transport} layer, and
  \item the \emph{synchronisation} layer.
  \end{enumerate} It proceeds as follows.
\begin{description}
\item[State layer.] Each component of a cell of $\ACB$ is simulated by a block of
  three adjacent cells in the state layer of $\limprod{\ACA}$.  More
  precisely, component $i$ (${0\leq i\leq n-1}$) of cell $z$ of $\ACB$
  is simulated by the block of three cells of $\limprod{\ACA}$ beginning
  at position ${3(nz+i)}$. This block is referred to as $\blc{z,i}$ in
  the sequel. In $\blc{z,i}$, the center cell stores the $i$th
  component of the cell $z$ of $\ACB$ and the two other are used to
  store temporarily the $i$th components of cell $z-1$ and $z+1$.
\item[Transport layer.] The role of the transport layer is precisely to bring states of
  $i$th components corresponding to cell $z-1$ and $z+1$ of $\ACA$ to
  the dedicated cells of $\limprod{\ACA}$ in $\blc{z,i}$. Then, the
  transition ${\locA(x_{z-1},x_{z},x_{z+1})}$ of the $i$th component of
  $\ACB$ can be simulated locally by $\limprod{\ACA}$ in
  $\blc{z,i}$. Transport is done in parallel for any $i$ and any
  $z$. To do this, the transport layer is made of a succession of
  particles (one every three cells), each one being able to carry a
  state of $\ACA$. Initially aligned with the center of blocks, the
  particles move in parallel according to a cycle of five steps:
  \begin{enumerate}
  \item move right by ${3n}$ cells and read the state seen on the state layer;
  \item move left by ${3n-1}$ cells and write the memorized state on
    the state layer;
  \item move left by ${3n+1}$ cells and read the state seen on the state layer;
  \item move left by ${3n-1}$ cells and write the memorized state on
    the state layer;
  \item move $1$ cell right and apply local rule $\locA$ on state
    layer at the current position;
  \end{enumerate}
\item[Synchronisation layer.] The role of the synchronisation layer is
  to orchestrate the cycle of particle moves and it must be able to do
  it for arbitrary large values of $n$ (simulating arbitrarily large
  Cartesian products of $\ACA$ is sufficient to simulate all products
  of $\ACA$). It contains a flag that can take one of the four
  indications 'left', 'right', 'read', 'write' and 'transition'. The
  flag is changed everywhere synchronously according to a cycle
  suitable to ensure that particles of the transport layer produce the
  cycle described above when they follow the instruction given by the
  flag.
\end{description}

We now describe in detail the synchronisation layer.  Denote by
$\flagseq{n}$ the flag sequence mentionned above in the simulation of
a product of $n$ copies of $\ACA$, and let $E$ be the set of flag
states.

\begin{thm}
  \label{thm:metronome}
  There is a CA $\ACC$ with a spreading state $\kappa$ and a map ${\pi
    : \alphC\rightarrow E}$ such that $\ACC$ is not
  $\facsacsimu$-universal, and, for any configuration
  $c\in\alphC^\ZZ$, one of the following property is true:
  \begin{description}
  \item[Cycle:] at each time in the orbit of $c$, all cells have the
    same image by $\pi$ and the sequence with time of this common
    image is periodic of period $\flagseq{n}$ for some $n$;
  \item[Frozen:] at each time in the orbit of $c$, all cells have the
    same image by $\pi$, but this common image remains constant after
    a certain time;
  \item[Error:] the spreading state appears at some time in the orbit
    of $c$.
  \end{description}
  Moreover $\ACC$ is such that there are configurations having the
  'cycle' property above producing period $\flagseq{n}$ for
  arbitrarily large $n$.
\end{thm}
\begin{pf}
  First, notice that flag changes in the sequence $\flagseq{n}$ are
  separated by a number of steps which is either constant (independant
  of $n$), or of the form ${3n+c}$ with $c$ a constant (we can suppose
  ${c\geq 0}$ without loss of generality). To simplify notations, we
  will suppose in this proof that $\flagseq{n}$ alternates between two
  values $0$ and $1$ every ${3n}$ steps. Adapting the proof for the
  real $\flagseq{n}$ is just a matter of adding a finite set of
  special states to deal with constants.
  
  The proof is based on a reversible solution $\ACB$ to the firing
  squad synchronization problem proposed by K.~Imai and K.~Morita: in
  \cite{ImaMor96}, they construct a reversible CA $\ACB$ with a
  subset of states $F$ (the firing states) such that for any $n$,
  there is a periodic configuration $c_n$ verifying\footnote{In
    \cite{ImaMor96}, the main concern is synchronisation of finite
    segments of cells surrounded by a quiescent state. To extend the
    property to infinite configurations, it is crucial that
    ``garbage'' (which must be conserved to ensure reversibility) do
    no spread outside the initial segment. The solution of K.~Imai and
    K.~Morita has precisely this property as it is explicitely
    mentionned in \cite{ImaMor96}.}:
  \begin{itemize}
  \item ${\globB^{3n}(c_n)\in F^\ZZ}$
  \item ${\globB^t(c_n)\in (\alphB\setminus F)^\ZZ}$ for all $t$,
    ${0\leq t<3n}$.
  \end{itemize}
  
  Without loss of generality, we can suppose that $\ACB$ and its
  inverse are syntactically represented with the same radius.  We now
  define a CA $\ACC_0$ of radius $r$, with states set
  ${\alphabe{\ACC_0}=\alphB\times \alphB\times\{0,1\}}$, and with
  transition function:
  \begin{align*}
    \local{\ACC_0}\bigl((a_{-r}, & a'_{-r},b_{-r}),\ldots,
    (a_{r},a'_{r},b_{r})\bigr) =\\
    &\begin{cases}
      \bigl(\locB(a_{-r},\ldots,a_{r}),\local{\ACB^{-1}}(a'_{-r'},
      \ldots,a'_{r'}),\chi(a_0,b_0)\bigr)
      &\text{if }b_0=1,\\
      \bigl(\local{\ACB^{-1}}(a_{-r},\ldots,a_{r}),\locB(a'_{-r},
      \ldots,a'_{r}),\chi(a'_0,b_0)\bigr) &\text{if }b_0=0,
    \end{cases}
  \end{align*}
  where ${\chi(a,b)}$ equals ${1-b}$ if ${a\in F}$ and $b$ else. Intuitively, on
  configurations where the third component is uniform equal to $b$, $\ACC_0$
  mimics $\ACB$ on the first component and $\ACB^{-1}$ on the second one if
  ${b=1}$ or the converse if ${b=0}$. Moreover, the value of $b$ is switched
  each time the component playing $\ACB$ encounters a firing state. Hence, if we
  choose for $\pi$ the projection on third component, $\ACC_0$ started from
  configurations $c_n$ has the property 'cycle' and produces the periodic
  sequence $\flagseq{n}$.

  We now enrich $\ACC_0$ with a spreading state which is produced each
  time one of the following local checking fails:
  \begin{itemize}
  \item the third component ${\{0,1\}}$ must be uniform;
  \item for the two first components, a state from $F$ (firing state)
    must always be surrounded by states from $F$ only;
  \item states from $F$ are forbidden on the second component if
    ${b=1}$ and states from $F$ in the first component are forbidden
    if ${b=0}$.
  \end{itemize}
  The third condition ensures that in the case of a 'cycle' regime
  (firing states appearing infinitely often), the period is equally
  divided between steps where ${b=0}$ and steps where ${b=1}$. To
  ensure that such a 'cycle' regime always produces an alternance of
  exactly $3n$ zeros and $3n$ ones, we add a component implementing a
  counter modulo $3$: the value of this component is incremented modulo
  3 at each step (whatever the context) and a spreading state is
  generated if a cell contains a firing state and the counter is not
  $0$ modulo $3$. Denote by $\ACC$ the CA obtained and consider any
  configuration $c$. If no spreading state appears in the orbit of
  $c$, then the third component is uniform. If it changes of state
  only a finite number of times, then we are in the 'frozen'
  regime. If there are infinitely many changes, it follows from the
  discussion above that the conditions of the 'cyclic' regime are
  fulfilled.

  To conclude the theorem, it remains to prove that $\ACC$ is not
  $\facsacsimu$-universal. Suppose for the sake of contradiction that it is and
  let $\ACU$ be any universal CA without spreading state and consider the set
  $X$ of configurations of $\ACC$ which is the support of the strong simulation
  of $\ACU$ ($\ACC$ strongly simulates $\ACU$ by
  theorem~\ref{thm:stronguniv}). $X$ cannot contain any occurrence of the
  spreading state (by lemma~\ref{lm:nospread}), it implies that all
  configurations of $X$ have a uniform third component. But, on such
  configuration, the dynamics of $\ACC$ is reversible. Hence $\ACU$ is
  reversible: this is a contradiction with its universality by
  theorem~\ref{thm:commonideal}.\qed
\end{pf}

The synchronization layer of limit products is exactly the automaton
$\ACC$ of the previous theorem, except that the spreading state of
$\ACC$ now becomes a global spreading state.  Before establishing the
main result of this section, we give more details concerning the state
layer and the transport layer of $\limprod{\ACA}$.

The state layer is made from state set ${\alphA\times\{L,C,R\}}$ where
$L$, $C$ and $R$ are states to identify explicitly the role of each
cell in each block $\blc{z,i}$: $C$ for the cell storing the $i$th
component of cell $z$ of $\ACA$ and $L$ and $R$ to temporarily store
states of $i$th component of cells $z-1$ and $z+1$ respectively.

The transport layer is made from state set ${\alphA\cup\{\bot\}}$
where $\bot$ is the state used to separate particle carrying a state
from $\alphA$.

So the states set of $\limprod{\ACA}$ is:
\[\underbrace{\alphA\times\{L,C,R\}}_{\text{state}}\ \times\
\underbrace{\alphA\cup\{\bot\}}_{\text{transport}}\ \times\
\underbrace{\alphC\setminus\{\kappa\}}_{\text{synchronization}}\ \cup\
\{\kappa\}.\] 

In addition to the behavior described above, $\limprod{\ACA}$ does the
following local checkings and generates the spreading state $\kappa$
if one of them fails:
\begin{itemize}
\item the second component of transport layer must be periodic of period ${LCR}$;
\item the transport layer must contain an alternance of one state from
  $\alphA$ and two states $\bot$;
\item when doing read and write operations, the particles of the
  transport layer must be aligned with the right type of state in the state layer:
  \begin{itemize}
  \item type $C$ when reading,
  \item type $R$ when writing at step $2$,
  \item type $L$ when writing at step $4$;
  \end{itemize}
\item when the synchronisation layer says 'transition', check that the
  particules are aligned with cell of type $C$ in the state layer.
\end{itemize}

All those checkings ensure the following property: if no spreading
state is generated and if the component layer produces a correct cycle
of instructions, then the behavior of the state layer is equivalent
to the behavior of some Cartesian product of $\ACA$ (up-to some
rescaling).

Before stating the main theorem, we establish a simple yet useful
lemma saying that if $\ACA$ simulates $\ACB$ with support $X$, then
everything $\ACA$ can simulate using a support included in $X$ can
also be simulated by $\ACB$.

\begin{lm}
  \label{lm:lift}
  Let $\simu$ be either $\sacsimu$ or $\facsacsimu$.  Let $\ACA$ and
  $\ACB$ be such that the simulation ${\ACA\simu\ACB}$ occurs on a
  support $X$ of configurations of $\ACB$. If $\ACB$ $\simu$-simulates
  $\ACC$ on a support included in $X$, then $\ACA$ $\simu$-simulates
  $\ACC$.
\end{lm}
\begin{pf}
  We consider the case where $\simu$ is $\sacsimu$.  By hypothesis, we
  have ${\bulk{\ACA}{\alpha}\sac_i\bulk{\ACB}{\beta_1}}$ on support
  $X$ and ${\bulk{\ACC}{\gamma}\sac_j\bulk{\ACB}{\beta_2}}$ on support
  ${Y\subseteq X}$. Now, let $m_\alpha$, $m_{\beta_1}$, $m_{\beta_2}$
  and $m_\gamma$ be the packing parameters of transforms $\alpha$,
  $\beta_1$, $\beta_2$ and $\gamma$ respectively. The injective maps
  $i$ and $j$ induce two injective maps $i_{\beta_2}$ and
  $j_{\beta_1}$ with the following domains and ranges:
  \begin{align*}
    i_{\beta_2} &: \alphA^{m_\alpha m_{\beta_2}}\rightarrow \alphB^{m_{\beta_1}m_{\beta_2}}\\
      j_{\beta_1} &: \alphC^{m_\gamma m_{\beta_1}}\rightarrow
      \alphB^{m_{\beta_1}m_{\beta_2}}
  \end{align*}
  Therefore ${\phi = i_{\beta_2}^{-1}\circ j_{\beta_1}}$ is a
  well-defined injective map from $\alphC^{m_\gamma m_{\beta_1}}$ into
  $\alphA^{m_\alpha m_{\beta_2}}$. Now define the transforms $\eta_a$
  and $\eta_c$ to be the composition of $\alpha$ and $\beta_2$, and of
  $\gamma$ and $\beta_1$ respectively. Then we have
  ${\bulk{\ACC}{\eta_c}\sac_\phi\bulk{\ACA}{\eta_a}}$.

  The extension of the previous reasoning to $\facsacsimu$ is
  straightforward.  \qed
\end{pf}

This lemma together with lemma~\ref{lm:nospread} is the key to a kind
of 'self-checking' simulation used in the construction of the limit
product (and re-used in section~\ref{sec:fam}). A 'self-checking'
simulation of $\ACB$ by $\ACA$ is standard simulation of $\ACB$ by
$\ACA$ on some support $X$ with the additional property that $\ACA$
'checks' locally on any configuration that it belongs to $X$, and
triggers some pathological behavior (typically a spreading state) in
case of check failure.  Hence any possible strong simulation of some
$\ACC$ by $\ACA$ is such that:
\begin{itemize}
\item either it has a support included in $X$ in which case $\ACB$ can
  also simulate $\ACC$ by lemma~\ref{lm:lift},
\item or it must contain some ${c\not\in X}$ in its support in which
  case a spreading state is generated and lemma~\ref{lm:nospread}
  gives some limitation on $\ACC$.
\end{itemize}

To show that a spreading state is generated in the second case above,
a crucial property is that the support of any simulation is by
definition always irreducible: if $u_1$ and $u_2$ are words occuring
in two configurations of the support, there exists a third
configuration of the support where $u_1$ and $u_2$ both appear (see
section 3.2 of \cite{bulking1} for a more detailed discussion on
supports of simulations).

We now state the main theorem of this section.

\begin{thm}
  \label{thm:limprod}
  For any $\ACA$, its limit product $\limprod{\ACA}$ is such that:
  \begin{itemize}
  \item ${\underbrace{\ACA\times\cdots\times\ACA}_{n}\facsacsimu\limprod{\ACA}}$ for all ${n\geq 1}$,
  \item $\limprod{\ACA}$ is $\facsacsimu$-universal if and only if
    $\ACA$ is $\facsacsimu$-universal.
  \end{itemize}
\end{thm}
\begin{pf}
  The first assertion follows from the construction of $\limprod{\ACA}$ and the
  detailed discussion above. Now suppose that $\limprod{\ACA}$ is
  $\facsacsimu$-universal and let $\ACU$ be any universal CA with no spreading
  state. By theorem~\ref{thm:stronguniv}, $\limprod{\ACA}$ strongly simulates
  $\ACU$: ${\ACU\facsac\bulk{\limprod{\ACA}}{\alpha}}$ for some geometrical
  transform $\alpha$. Let $X$ denote the support of the simulation. By choice of
  $\ACU$, the spreading state $\kappa$ cannot appear in any orbit of any
  configuration of $X$ (by lemma~\ref{lm:nospread}). We deduce from
  theorem~\ref{thm:metronome} that the synchronization component is in the same
  regime (either 'cycle' for a fixed value $n$ or 'frozen') for all the
  configurations of $X$ because otherwise, we could construct a configuration in
  $X$ producing a spreading state (by irreducibility of $X$).

  In the case where all configurations are in the frozen regime, the
  flag of the synchronization layer becomes constant after some time
  $t_0$, so the transport layer has the behavior of a translation (or
  identity) and the state layer remains constant. $t_0$ is identical
  for all configurations of $X$ (because otherwise, we could once more
  combine two configurations to produce a spreading state, by
  irreducibility of $X$). Then, consider a CA $\ACU_+$ with state set
  ${\alphabe{\ACU}\times\{0,\ldots,t_0\}}$ which has the following
  behavior:
  \begin{itemize}
  \item the second component is decreased by one until it reaches $0$;
  \item on the first component, the local rule of $\ACU$ is applied,
    but only if the second component is $0$.
  \end{itemize}
  Since $\ACU$ is $\facsacsimu$-universal, it can strongly simulate
  $\ACU_+$ (by theorem~\ref{thm:stronguniv}): precisely,
  ${\ACU_+\facsac\bulk{\ACU}{m,1,t,z}}$. Consider the set $Y$ of
  configurations of $\ACU$ corresponding via simulation to the set of
  configurations of $\ACU_+$ uniformly equal to $t_0$ on the second
  component. Denote by ${X_Y\subseteq X}$ the corresponding set of
  configurations of $\limprod{\ACA}$. By choice of $\ACU_+$, we know
  that $\ACU$ simulates itself on the set of configurations
  ${\glob{\ACU}^{tt_0}(Y)}$. This implies that for some ${t'\geq
    t_0}$, $\limprod{\ACA}$ can simulate $U$ using as support the set
  of configurations ${\glob{\limprod{\ACA}}^{t'}(X_Y)}$. By hypothesis,
  starting from such configurations, $\limprod{\ACA}$ has a behavior
  of translation or identity on the state and transport layers. Since
  the synchronizing component evolves independently of the others, we
  deduce by lemma~\ref{lm:lift} that there is some CA $\ACB$ which is
  a product of translations (corresponding to state and transport
  layers) such that ${\ACB\times\ACC}$ simulates $\ACU$: this is a
  contradiction by theorem~\ref{thm:notuniv} since neither $\ACB$
  (theorem~\ref{thm:shishi}), nor $\ACC$
  (theorem~\ref{thm:metronome}) is universal.

  Hence, we are necessarily in the case where the synchronization
  layers produce a valid cycle. Since no spreading state can be
  generated in the orbit of any configuration of $X$, the state layer
  always behaves like a Cartesian product of $n$ copies of $\ACA$. The
  value of $n$ is in fact common to all configurations of $X$ (as
  shown above), so we deduce by lemma~\ref{lm:lift} that
  ${\underbrace{\ACA\times\cdots\times\ACA}_n}$ simulates $\ACU$ and
  $\ACA$ is therefore universal by theorem~\ref{thm:notuniv}. \qed
\end{pf}

Of course, we can consider $\limprod{\ACA}$ itself as a new candidate
for taking its finite Cartesian products and applying the limit
product construction. In fact, the process can be repeated forever
with the guarantee that no CA ever produced in this chain will be
universal, provided the initial CA is not. However, there is no reason
why this infinite chain should be strictly increasing. In particular,
even if
\[\ACA\simu\ACA\times\ACA\simu\ACA\times\ACA\times\ACA\simu\cdots\]
is a strictly increasing chain, it might be the case that
${\limprod{\ACA}}$ is equivalent to
${\limprod{\ACA}\times\limprod{\ACA}}$. Therefore we have only proven
that one of the following properties is true:
\begin{itemize}
\item there is a strictly increasing chain of length
  $\omega\times\omega$ in the quasi-order ${(AC,\facsacsimu)}$,
\item for any non-universal CA $\ACA$, there is a non-universal CA
  $\ACB$ such that ${\ACA\facsacsimu\ACB}$ and $\ACB\times\ACB$ is
  equivalent to $\ACB$.
\end{itemize}

\subsection{Sub-Families of Cellular Automata}
\label{sec:fam}

Theorem~\ref{thm:bfacts} shows that any equivalence class in any
quasi-order contains some CA with radius $1$. This fact is a direct
consequence of a well-known transformation of CA with large radius
into CA of smaller radius with more states (this transformation is
called 'higher block presentation' in symbolic dynamics,
see~\cite{LindMarcus}).

It is also sometimes invoked in the literature that considering CA
with two states only is not restrictive since there is a converse
transformation that transforms a CA with many states into a CA with
only two states but a larger radius. However, the situation is not
similar to that of radius reduction since there are equivalence
classes with no $2$-states CA: \textit{e.g.}  $\zee{p}$ for any prime
${p\not=2}$ as shown by theorem~\ref{thm:mazrapadd} and
lemma~\ref{lm:manulem}. Note that the same is true for any fixed state
set of cardinal $n$: the equivalence class of $\zee{p}$ contains no
such CA provided $p$ is prime and does not divide $n$.

Hence, this transformation introduces a bias: the transformed CA may
be inequivalent to the original one. Meanwhile, we know that CA with
two states can be as powerful as CA in general since there are
universal CA with two states only \cite{banks,nicolasFCT} (for
simulation relations $\sacsimu$ and $\facsacsimu$). More precisely, as
we will see below, the transformation applied to a universal CA always
yields a universal CA because the transformed CA simulates the
original one. Since the original and the transformed CA are not always
in the same equivalence class, one question that naturally arises is:
what CA can be simulated by the transformed CA but not by the original
one?  Although it provides only partial answers, this section is
devoted to that kind of questions, for CA with two states and for
other families.

Formally, given a familly $\fami$ of CA, we say that a map
${\cod:AC\rightarrow \fami}$ is a $\simu$-encoding of CA into familly
$\fami$ if
\[\forall\ACA, \ACA\simu\cod(\ACA).\]
We will only consider simulation relations $\sacsimu$ and $\facsacsimu$ in the
sequel, thus an encoding into $\fami$ implies that there are universal CA in
$\fami$. A trivial example of such an encoding is given by ${\fami=\{\ACU\}}$
where $\ACU$ is a universal CA and $\phi$ is the function mapping any CA to
$\ACU$.  We are interested in using this notion of encoding with families which
are more 'representative' of the diversity of behaviors in the whole set of CA.
To express this we introduce the following notion of \emph{faithfulness}.

Given a $\simu$-encoding ${\cod:AC\rightarrow \fami}$ and a set $E$ of
CA, we say that $\cod$ is \emph{faithful} for $E$ if:
\[\forall\ACB\in E: \ACB\simu\ACA\iff\ACB\simu\cod(\ACA).\]
An encoding is faithful for $E$ if the original CA and its image by
the encoding simulate exactly the same CA in $E$. So, to give some
evidence that a family $\fami$ is 'representative' of CA in general,
we can exhibit an encoding of CA into $\fami$ which is faithful for a
set $E$ of CA as large as possible. When $E$ is the whole set of CA,
the faithfulness implies that there is a CA of family $\fami$ in any
equivalence class: this is the case for CA with radius $1$.\\

The next theorem gives four encodings which are faithful for $U$, the
set of $\facsacsimu$-universal CA. The families corresponding to these
encodings were already defined in this paper except one: captive CA.

Captive CA were introduced in \cite{Theyssier04} and are defined by a
simple restriction on the transition rule. A CA $\ACA$, with state set
$\alphA$, radius $r$ and local rule $\locA$ is captive if:
\[\forall a_{-r},\ldots,a_r\in\alphA:
\locA(a_{-r},\ldots,a_r)\in\{a_{-r},\ldots,a_r\}.\]

In the following theorem, encodings are different but their
faithfulness rely on the same idea of 'self-checking' simulation
explained above which uses lemma~\ref{lm:nospread} and
lemma~\ref{lm:lift}.

\begin{thm}
  \label{thm:encodings}
  Let $\simu$ be $\sacsimu$ or $\facsacsimu$.  For any family of CA
  below, there is a $\simu$-encoding from CA into $\fami$ wich is
  faithful for the set $U$ of $\facsacsimu$-universal CA:
  \begin{itemize}
  \item CA with two states,
  \item CA in $\equipt$,
  \item CA in $\sensi$, 
  \item captive CA.
  \end{itemize}
\end{thm}
\begin{pf}
  To describe the encoding for each family, we suppose $\ACA$ is a CA
  with state set ${\alphA=\{a_1,\ldots,a_n\}}$, with radius $r$ and
  location rule $\locA$.
  \begin{description}
  \item[$2$-states CA.] Let $m$ be an integer large enough and $\psi$
    be an injective map from $\alphA$ to ${\{0,1\}^m}$ such that no
    word $\psi(a)$ contains an occurrence of $11$. Now define the
    injective map ${i:\alphA\rightarrow\{0,1\}^{m+4}}$ by
    ${i(a)=0110\psi(a)}$. Let ${r'=(r+1)(m+4)}$. ${\cod(\ACA)}$ is a
    CA of radius $r'$ and state set ${\{0,1\}}$ defined as follows:
    \begin{itemize}
    \item on the set $X$ of configurations made of infinite
      concatenations of words from $i(\alphA)$, $\cod(\ACA)$ is
      isomorphic to $\ACA$ so that ${\ACA\sac_i\cod(\ACA)}$;
    \item everywhere else, $\cod(\ACA)$ generates a $1$.
    \end{itemize}
    The map $\cod$ is thus an encoding of CA into $2$-states CA. Now
    suppose that $\cod(\ACA)$ is universal and let $\ACU$ be a
    universal CA with two states and no spreading state which is
    strongly simulated by $\cod(\ACA)$ on support $Y$
    (theorem~\ref{thm:stronguniv}). If there is some ${y\in Y}$ with
    ${y\not\in X}$ then
    \begin{itemize}
    \item either there are two occurrences of $0110$ in $y$ which are
      not correctly spaced,
    \item or there is a word ${0110u0110}$ occurring in $y$ with
      ${u\not\in\psi(\alphA)}$.
    \end{itemize}
    In any case, the image of $y$ will contain an occurrence of $111$
    (because the above error must be seen by at least three consecutive
    cells) and $1$'s will propagate like a spreading state which is
    impossible by lemma~\ref{lm:nospread} because otherwise
    $\bulk{\cod(\ACA)}{3,1,1,0}$ could simulate $\ACU$ on a support
    where it possesses the spreading state $111$. So ${Y\subseteq X}$
    and lemma~\ref{lm:lift} shows that $\ACA$ simulates $\ACU$. Hence
    $\ACA$ is universal if and only if $\cod(\ACA)$ is.

  \item[Captive CA.] The encoding technique for captive CA is very
    similar and already appeared in a non-faithful form in
    \cite{Theyssier04}. Let $u$ be the word ${a_1\cdots a_n}$, let
    ${\#}$ be a state not in $\alphA$ and denote
    ${Q=\alphA\cup\{\#\}}$. We define the injective map
    ${i:\alphA\rightarrow Q^{n+3}}$ by ${i(a) = \#u\#a}$. We then
    define $\cod(\ACA)$ in a way similar to the case above. Its radius
    is ${r'=(r+1)(n+3)}$, its state set is $Q$ and its local rule is
    such that:
    \begin{itemize}
    \item on the set $X$ of configurations made of infinite
      concatenations of words from $i(\alphA)$, $\cod(\ACA)$ is
      isomorphic to $\ACA$ so that ${\ACA\sac_i\cod(\ACA)}$;
    \item everywhere else, $\cod(\ACA)$ take as new state the maximum
      of its neighbors for some fixed ordering of $Q$ such that $\#$
      is the maximum.
    \end{itemize}    
    First, $\cod(\ACA)$ is captive and $\cod$ defines an encoding of
    CA into captive CA. Second, notice that for any support of
    simulation $Y$ of $\cod(\ACA)$, if there is some ${y\in Y}$ with
    ${y\not\in X}$ then, by irreducibility of $Y$, either there is
    ${y'\in Y}$ with ${y'\not\in X}$ and $y'$ contains a $\#$, or $\#$
    never appears in $Y$. In the second case, $\cod(\ACA)$ always
    applies a $\max$ as local rule and therefore possesses a spreading
    state when restricted to $Y$. In the first case, consider the
    configuration $y'$ and ${z\in\ZZ}$ such that positions ${z}$ and
    ${z+1}$ both see a $\#$ in their neighborhood and a local pattern
    not in $X$ (such a $z$ must exist by choice of $y'$ and definition
    of $X$). Then $\cod(\ACA)(y')$ contains the pattern $\#\#$ which
    is spreading by definition of $\cod(\ACA)$. In any case we can
    apply the usual reasoning with lemma~\ref{lm:nospread} and
    lemma~\ref{lm:lift}: any CA without spreading state strongly
    simulated by $\cod(\ACA)$ is also simulated by $\ACA$. So the
    encoding $\cod$ is faithful for universal CA.
  \item[$\equipt$ and $\sensi$.] For $\equipt$, the encoding is
    simple: $\cod(\ACA)$ is just $\ACA$ with an additional state
    $\kappa$ which is spreading. The resulting CA $\cod(\ACA)$ is
    always in $\equipt$ since $\kappa^{2r}$ is a blocking word (see
    \cite{Kurka97}). Lemma~\ref{lm:nospread} is then sufficient to
    prove that it is an encoding from AC to $\sensi$ which is faithful
    for universal CA.

    For $\sensi$, the proof is even simpler:
    ${\cod(\ACA)=\ACA\times\sigma_1\times\sigma_{-1}}$ is always in
    $\sensi$ since $\sigma_1\times\sigma_{-1}\in\sensi$ and an
    equicontinuous point in a Cartesian product induce equicontinuous
    points for each component.  Theorem~\ref{thm:notuniv} concludes
    for the faithfulness.
  \end{description}
  \qed
\end{pf}

These encodings allow to transport some properties of general CA
concerning the top of quasi-orders into order structures induced by
each family\footnote{A stronger result concerning captive CA appears
  in \cite{Theyssier05}: $\sacsimu$-universality is undecidable even if
  we restrict to captive CA with a fixed (but sufficiently large)
  radius.}.

\begin{cor}
  Let $\simu$ be $\sacsimu$ or $\facsacsimu$ and let $\fami$ be a
  family of CA among: CA with two states, $\equipt$, $\sensi$, captive
  CA.  Then we have the following properties:
  \begin{itemize}
  \item the set of $\simu$-universal CA in $\fami$ is not co-r.e.
  \item for any non-universal $\ACA\in\fami$, there is a non universal
    $\ACB\in\fami$ with ${\ACA\simu\ACB}$ but ${\ACB\cancel\simu\ACA}$.
  \end{itemize}
\end{cor}
\begin{pf}
  The first property is a direct corollary of
  theorem~\ref{thm:encodings} and~\ref{thm:indeciuni} by definition of
  faithful encodings.

  For the second property, consider the encoding $\phi$ established in
  theorem~\ref{thm:encodings} and let ${\ACA\in\fami}$ be any
  non-universal CA. By theorem~\ref{thm:notuniv}, there is some
  non-universal CA $\ACB$ such that ${\ACA\simu\ACB}$ but
  ${\ACB\cancel\simu\ACA}$. By faithfulness of $\phi$,
  $\phi(\ACB)\in\fami$ is not universal and by the definition of
  encoding it simulates $\ACA$ without being simulated by $\ACA$.\qed
\end{pf}

The families considered above induce structures sharing some
properties with the general quasi-orders 'near the top'. However, the
complete characterisation of equivalence classes occupied by some CA
of these families is more challenging.

\begin{openpb}
  What are the quivalence classes of the simulation quasi-orders
  containing a $2$-states CA? a captive CA? a CA from $\equipt$? a CA
  from $\sensi$?
\end{openpb}

\section{Summary of results}

Figures~\ref{fig:injbulking}, \ref{fig:surjbulking} and \ref{fig:mixbulking}
hereafter give a summary of results and open problems concerning each of the three
quasi-orders studied in the paper.





\newcommand\class[2]{\draw[fill=white] #1 circle (.47cm);\draw #1 node {#2};}

\newcommand\coolclass[2]{\draw #1 node[ellipse,fill=white] {#2};}

\newcommand\simlink[2]{\draw #1 -- #2;}

\newcommand\dontknow[2]{\draw #1 node[ellipse,draw,fill=gray!20] {#2};}

\newcommand\thmref[3]{\path #1 -- #2 node[sloped,above,pos=0.40] {#3};}

\newcommand\lebas{(1,0)}
\newcommand\univcls{(3,15)}
\newcommand\runiv{(-.8,7)}
\newcommand\topsurj{(-1.8,10)}
\newcommand\topexp{(-3.3,5.3)}
\newcommand\zpz{(-3.5,2)}
\newcommand\zdz{(-2,2)}
\newcommand\nilpo{(7.5,2)}
\newcommand\iden{(2.4,2)}
\newcommand\sudt{(1.5,3.2)}
\newcommand\sudq{(0,3.2)}
\newcommand\sudtq{(.4,4.6)}
\newcommand\sudtc{(-1.1,4.6)}
\newcommand\turuniv{(6,6)}

\newcommand\surjgauche{(-6,1)}
\newcommand\surjdroite{(5,2)}
\newcommand\revgauche{(-2.2,4.3)}
\newcommand\revdroite{(4,2)}
\newcommand\expgauche{(-4.7,1.5)}
\newcommand\expdroite{(-1,2)}
\newcommand\expbas{(-2,.7)}

\thispagestyle{empty}

\begin{figure}
  \hskip -2cm\hbox{%
  \begin{tikzpicture}
    \draw[bottom color=yellow!80,top color=yellow!10] \lebas..controls
    \surjgauche..\topsurj..controls \surjdroite..\lebas;
    \draw[bottom color=orange!80,top color=orange!10] \lebas..controls
    \revgauche..\runiv..controls \revdroite..\lebas;
    \draw[bottom color=red!80,top color=red!10] \expbas..controls
    \expgauche..\topexp..controls \expdroite..\expbas;

    \draw[dotted] \zpz++(-1.5,0)--(9.5,2);%

    \thmref{(8.5,2)}{(9.5,2)}{\small level 1}
    \thmref{\topsurj}{\surjgauche}{\tiny Surjective ideal, Thm. \ref{thm:commonideal}}%
    \thmref{\runiv}{\revdroite}{\tiny Reversible ideal, Thm. \ref{thm:commonideal}}%
    \thmref{\sudtc++(0,.5)}{+(1,0)}{\tiny Thm. \ref{thm:noshishi}}%
    \thmref{\univcls++(-.5,.5)}{+(1,0)}{\tiny Thm. \ref{thm:stronguniv}}%
    \thmref{\topexp}{\expgauche}{\tiny Thm. \ref{thm:tdideals}}%
    \thmref{\topexp}{\expdroite}{\tiny Expansive ideal}%
    \coolclass{\univcls++(0,-2)}{\tiny Thm. \ref{thm:notuniv}}%
    \thmref{\lebas}{\zpz}{\tiny Thm. \ref{thm:zpz}}%
    \thmref{\runiv++(0,.5)}{+(-1,0)}{\tiny Thm. \ref{thm:revuniv}}%
    \thmref{\iden}{\lebas}{\tiny Thm. \ref{thm:nilper}}%
    \thmref{\lebas}{\nilpo}{\tiny Thm. \ref{thm:nilper}}%
    \thmref{\turuniv++(0,1)}{+(0,2.5)}{\tiny Thm. \ref{thm:turingheads}}%

    \dontknow{\topsurj}{\small Open Pb. \ref{open:surjideal}}%
    \dontknow{\topexp}{?}%
    \dontknow{(5.2,2)}{\small Open Pb. \ref{open:abovenilpo}}

    \simlink{\lebas}{\zpz}%
    \simlink{\lebas}{\zdz}%
    \simlink{\lebas}{\iden}%
    \simlink{\lebas}{\nilpo}%
    \simlink{\iden}{\sudt}%
    \simlink{\iden}{\sudq}%
    \simlink{\sudt}{\sudtq}%
    \simlink{\sudt}{\sudtc}%
    \simlink{\sudq}{\sudtq}%
    \simlink{\sudq}{\sudtc}%
    \foreach \i in {-3,-1,1,3} \draw[dashed] \univcls--+(\i,-2.5);%
    \draw[dashed] \nilpo--+(-2,2);%
    \draw[dashed] \nilpo--+(2,2);%
    \simlink{\turuniv}{+(0,1)}; \draw[dashed] \turuniv++(0,1)--+(0,2);%

    \coolclass{\turuniv}{$\mathcal A$ (Turing-universal)}%
    \coolclass{\turuniv+(0,1)}{$\mathcal A\times\mathcal A$}%
    \coolclass{\turuniv+(0,3.5)}{$\mathcal A\times\cdots\times\mathcal
      A$}%
    \class{\lebas}{$\bot$}%
    \class{\univcls}{U}%
    \class{\runiv}{RU}%
    \class{\zpz}{\small$\mathbb{Z}/p\mathbb{Z}$}%
    \class{\zdz}{\small$\mathbb{Z}/2\mathbb{Z}$}%
    \class{\nilpo}{NIL}%
    \class{\iden}{$\sigma$}%
    \class{\sudt}{\tiny $\sigma^{1,2,3}$}%
    \class{\sudq}{\tiny $\sigma^{1,2,4}$}%
    \class{\sudtq}{\tiny $\sigma^{1,2,3,4}$}%
    \class{\sudtc}{\tiny $\sigma^{1,2,3,5}$}%
  \end{tikzpicture}}
  
  \caption{Injective bulking (quasi-order $\sacsimu$)}
\label{fig:injbulking}
\end{figure}

\begin{figure}
  \hskip -2cm\hbox{%
  \begin{tikzpicture}
    \begin{scope}
      \draw[bottom color=yellow!80,top color=yellow!10] \lebas..controls
      \surjgauche..\topsurj..controls \surjdroite..\lebas;
      \draw[bottom color=orange!80,top color=orange!10] \lebas..controls
      \revgauche..\runiv..controls \revdroite..\lebas;
      \draw[fill=gray!20] \expbas..controls \expgauche..\topexp..controls
      \expdroite..\expbas;

      \draw[dotted] \zpz++(-1.5,0)--(9.5,2);%

      \thmref{(8.5,2)}{(9.5,2)}{\small level 1}
      \thmref{\topsurj}{\surjgauche}{\tiny Surjective ideal, Thm. \ref{thm:commonideal}}%
      \thmref{\runiv}{\revdroite}{\tiny Reversible ideal, Thm. \ref{thm:commonideal}}%
      \thmref{\sudtc++(0,.5)}{+(1,0)}{\tiny Thm. \ref{thm:noshishi}}%
      \thmref{\univcls++(-.5,.5)}{+(1,0)}{\tiny Thm. \ref{thm:stronguniv}}%
      \thmref{\topexp}{\expgauche}{\small Open Pb. \ref{open:expansif}}%
      \thmref{\topexp}{\expdroite}{\tiny Set of expansive CA}%
      \coolclass{\univcls++(0,-2)}{\tiny Thm. \ref{thm:notuniv}}%
      \thmref{\lebas}{\zpz}{\tiny Thm. \ref{thm:zpz}}%
      \thmref{\runiv++(0,.5)}{+(-1,0)}{\tiny Thm. \ref{thm:revuniv}}%
      \thmref{\iden}{\lebas}{\tiny Thm. \ref{thm:nilper}}%
      \thmref{\lebas}{\nilpo}{\tiny Thm. \ref{thm:nilper}}%
      \thmref{\turuniv++(0,1)}{+(0,2.5)}{\tiny Thm. \ref{thm:turingheads}}%
      \thmref{\turuniv++(0,3.5)}{+(0,2.5)}{\tiny Thm. \ref{thm:limprod}}%
      \thmref{(5,2)}{+(1,0)}{\tiny empty set, Thm. \ref{thm:abovenilpo}}%

      \dontknow{\topsurj}{\small Open Pb. \ref{open:surjideal}}%
      \dontknow{\topexp}{?}

      \simlink{\lebas}{\zpz}%
      \simlink{\lebas}{\zdz}%
      \simlink{\lebas}{\iden}%
      \simlink{\lebas}{\nilpo}%
      \simlink{\iden}{\sudt}%
      \simlink{\iden}{\sudq}%
      \simlink{\sudt}{\sudtq}%
      \simlink{\sudt}{\sudtc}%
      \simlink{\sudq}{\sudtq}%
      \simlink{\sudq}{\sudtc}%
      \foreach \i in {-3,-1,1,3} \draw[dashed] \univcls--+(\i,-2.5);%
      \draw[dashed] \nilpo--+(-2,2);%
      \draw[dashed] \nilpo--+(2,2);%
      \simlink{\turuniv}{+(0,1)}; %
      \draw[dashed] \turuniv++(0,1)--+(0,2.5);%
      \draw[dashed] \turuniv++(0,3.5)--+(0,2.5);%

      \coolclass{\turuniv}{$\mathcal A$ (Turing-universal)}%
      \coolclass{\turuniv+(0,1)}{$\mathcal A\times\mathcal A$}%
      \coolclass{\turuniv+(0,3.5)}{$\mathcal A\times\cdots\times\mathcal A$}%
      \coolclass{\turuniv+(0,6)}{${\mathcal A}_\infty$}%
      \class{\lebas}{$\bot$}%
      \class{\univcls}{U}%
      \class{\runiv}{RU}%
      \class{\zpz}{\small$\mathbb{Z}/p\mathbb{Z}$}%
      \class{\zdz}{\small$\mathbb{Z}/2\mathbb{Z}$}%
      \class{\nilpo}{NIL}%
      \class{\iden}{$\sigma$}%
      \class{\sudt}{\tiny $\sigma^{1,2,3}$}%
      \class{\sudq}{\tiny $\sigma^{1,2,4}$}%
      \class{\sudtq}{\tiny $\sigma^{1,2,3,4}$}%
      \class{\sudtc}{\tiny $\sigma^{1,2,3,5}$}%
    \end{scope}
  \end{tikzpicture}}
  
  \caption{Mixed bulking (quasi-order $\facsacsimu$)}
\label{fig:mixbulking}
\end{figure}

\begin{figure}
  \hskip -2cm\hbox{%
  \begin{tikzpicture}[xshift=-4cm]

    \newcommand\slstop{(-2.2,13)}
    \newcommand\slsdroite{(11.3,1)}
    \newcommand\slsgauche{(-7,1)}
    \draw[bottom color=green!30,top color=green!5] \lebas..controls
    \slsgauche..\slstop..controls \slsdroite..\lebas;

    \draw[bottom color=yellow!80,top color=yellow!10] \lebas..controls
    \surjgauche..\topsurj..controls \surjdroite..\lebas;
    \draw[bottom color=orange!80,top color=orange!10] \lebas..controls
    \revgauche..\runiv..controls \revdroite..\lebas;
    \draw[fill=gray!20] \expbas..controls
    \expgauche..\topexp..controls \expdroite..\expbas;

    \newcommand\tdgauche{(1.5,2.5)}
    \newcommand\tddroite{(9.3,1)}
    \newcommand\tdtop{(6.5,6.5)}
    \draw[bottom color=blue!80,top color=blue!10, fill opacity=.5] \lebas..controls
    \tdgauche..\tdtop..controls \tddroite..\lebas;

    \draw[dotted] \zpz++(-1.5,0)--(9.5,2);%

    \thmref{\tdtop}{\tdgauche}{\tiny $T2$ ideal, Thm. \ref{thm:tdideals}}
    \thmref{\slstop}{\slsdroite}{\tiny ideal of CA with a sofic limit set, Thm. \ref{thm:soficideal}}
    \thmref{(9,2)}{(9.5,2)}{\small level 1}
    \thmref{\topsurj}{\surjgauche}{\tiny Surjective ideal, Thm. \ref{thm:commonideal}}%
    \thmref{\runiv}{\revdroite}{\tiny Reversible ideal, Thm. \ref{thm:commonideal}}%
    \thmref{\sudtc++(0,.5)}{+(1,0)}{\tiny Thm. \ref{thm:noshishi}}%
    \thmref{\topexp}{\expgauche}{\small Open Pb. \ref{open:expansif}}%
    \thmref{\topexp}{\expdroite}{\tiny Set of expansive CA}%
    \thmref{\lebas}{\zpz}{\tiny Thm. \ref{thm:zpz}}%
    \thmref{\iden}{\lebas}{\tiny Thm. \ref{thm:nilper}}%
    \thmref{\lebas}{\nilpo}{\tiny Thm. \ref{thm:nilper}}%
    \thmref{(5,2)}{+(1,0)}{\tiny empty set, Thm. \ref{thm:abovenilpo}}%

    \dontknow{\topsurj}{\small Open Pb. \ref{open:surjideal}}%
    \dontknow{\topexp}{?}%
    \dontknow{\tdtop}{?}%
    \dontknow{\slstop}{?}%
    \dontknow{\runiv}{?}%
    \dontknow{\univcls}{\small Open Pb. \ref{open:surjuniv}}%

    \simlink{\lebas}{\zpz}%
    \simlink{\lebas}{\zdz}%
    \simlink{\lebas}{\iden}%
    \simlink{\lebas}{\nilpo}%
    \simlink{\iden}{\sudt}%
    \simlink{\iden}{\sudq}%
    \simlink{\sudt}{\sudtq}%
    \simlink{\sudt}{\sudtc}%
    \simlink{\sudq}{\sudtq}%
    \simlink{\sudq}{\sudtc}%
    \draw[dashed] \nilpo--+(-2,2);%
    \draw[dashed] \nilpo--+(2,2);%

    \class{\lebas}{$\bot$}%
    \class{\zpz}{\small$\mathbb{Z}/p\mathbb{Z}$}%
    \class{\zdz}{\small$\mathbb{Z}/2\mathbb{Z}$}%
    \class{\nilpo}{ NIL}%
    \class{\iden}{$\sigma$}%
    \class{\sudt}{\tiny $\sigma^{1,2,3}$}%
    \class{\sudq}{\tiny $\sigma^{1,2,4}$}%
    \class{\sudtq}{\tiny $\sigma^{1,2,3,4}$}%
    \class{\sudtc}{\tiny $\sigma^{1,2,3,5}$}%
  \end{tikzpicture}}
  
  \caption{Surjective bulking (quasi-order $\facsimu$)}
\label{fig:surjbulking}
\end{figure}


\newpage
\bibliographystyle{elsart-num-sort}
\bibliography{bulk2}

\end{document}